\documentclass[journal]{IEEEtran}

\usepackage{xcolor}

\usepackage{url}
\usepackage[acronym]{glossaries}
\newacronym{thz}{THz}{Terahertz}
\newacronym{etsi}{ETSI}{European Telecommunications Standards Institute}
\newacronym{isg}{ISG}{Industry Specification Group}
\newacronym{itu}{ITU}{International Telecommunication Union}
\newacronym{wrc}{WRC} {World Radiocommunication Conference}
\newacronym{eess}{EESS}{Earth Exploration Satellite Services}
\newacronym{ras}{RAS}{Radio Astronomy Services}
\newacronym{ieee}{IEEE}{Institute of Electrical and Electronics Engineers}
\newacronym{3gpp}{3GPP}{3rd Generation Partnership Project}
\newacronym{gr}{GR}{Group Report}
\newacronym{xr}{XR}{eXtended Reality}
\newacronym{vr}{VR}{Augmented Reality}
\newacronym{ar}{AR}{Virtual Reality}
\newacronym{mr}{MR}{Mixed Reality}
\newacronym{d2d}{D2D}{Device-to-Device}
\newacronym{iot}{IoT}{Internet of Things}
\newacronym{dt}{DT}{Digital Twin}
\newacronym{kpi}{KPI}{Key Performance Indicator}
\newacronym{mmwave}{mmWave}{millimeter wave}
\newacronym{ai}{AI}{Artificial Intelligence}
\newacronym{mimo}{MIMO}{Multiple Input Multiple Output}
\newacronym{los}{LOS}{Line Of Sight}
\newacronym{nlos}{NLOS}{Non Line Of Sight}
\newacronym{ris}{RIS}{Reconfigurable Intelligent Surface}
\newacronym{em}{EM}{electromagnetic}
\newacronym{isac}{ISAC}{Integrated Sensing and Communication}
\newacronym{o2i}{O2I}{Outdoor-to-Indoor}
\newacronym{ue}{UE}{User Equipment}
\newacronym{llm}{LLM}{Large Language Model}
\newacronym{ml}{ML}{Machine Learning}
\newacronym{6g}{6G}{sixth generation}
\newacronym{emf}{EMF}{Electromagnetic Field}
\newacronym{rf}{RF}{Radio Frequency}
\newacronym{pa}{PA}{Power Amplifier}
\newacronym{papr}{PAPR}{Peak-to-Average Power Ratio}
\newacronym{dpd}{DPD}{Digital Pre-Distortion}
\newacronym{ofdm}{OFDM}{Orthogonal Frequency Division Multiplexing}
\newacronym{pd}{PD}{Predistortion}
\newacronym{ber}{BER}{Bit Error Rate}
\newacronym{evm}{EVM}{Error Vector Magnitude}
\newacronym{lo}{LO}{Local Oscillator}
\newacronym{dut}{DUT}{Device Under Test}
\newacronym{mmic}{MMIC}{Monolithic Microwave Integrated Circuit}
\newacronym{cw}{CW}{Continuous Wave}
\newacronym{awg}{AWG}{Arbitrary Waveform Generator}
\newacronym{awgn}{AWGN}{Additive White Gaussian Noise}
\newacronym{ibo}{IBO}{Input Back-off}
\newacronym{rms}{RMS}{Root Mean Square}
\newacronym{phy}{PHY}{Physical}
\newacronym{amam}{AM-AM}{Amplitude-to- Amplitude Modulation}

\newacronym{ampm}{AM-PM}{Amplitude-to-Phase Modulation}

\newacronym{vna}{VNA}{Vector Network Analyzer}
\usepackage{amsmath}
\usepackage{amssymb}
\usepackage{algorithmic}
\usepackage{float}
\usepackage[font=small]{caption}
\usepackage{subcaption}
\usepackage{tikz}
\usepackage{pgfplots}
\usetikzlibrary{shapes,shadows,arrows.meta,intersections,decorations.pathmorphing,decorations.markings}

\usepackage{graphicx} % Required for inserting images
\usepackage{epstopdf}
\newcommand{\norm}[1]{\left\lVert#1\right\rVert}
\title{Sub-THz Power Amplifiers: Measurements,
Behavioral Modeling and Predistortion Algorithms}
	\author{Lutfi Samara$^{1,2}$, Simon Haussmann$^{3}$, Erind Tufa$^{4}$, Antonio Alberto D'Amico$^{4,5}$, Tommaso Zugno$^{1}$, Ingmar Kallfass$^{3}$, Thomas Kürner$^{2}$\\
		$^{1}$Munich Research Center, Huawei Technologies, Munich, Germany\\
		$^{2}$Technische Universität Braunschweig, Braunschweig, Germany\\
        $^{3}$University of Stuttgart, Inst. of Robust Power Semiconductor Systems, Stuttgart, Germany\\
        $^{4}$Dipartimento di Ingegneria dell’Informazione, University of Pisa, Pisa, Italy\\
        $^{5}$National Inter-University Consortium for Telecommunications (CNIT), Parma, Italy\\
		Emails: \{lutfi.samara,tommaso.zugno\}@huawei.com,\{simon.haussmann,ingmar.kallfass\}@ilh.uni-stuttgart.de,erind.tufa@phd.unipi.it, antonio.damico@unipi.it, t.kuerner@tu-braunschweig.de \thanks{This work has been performed in part in the framework
of the HORIZON-JU-SNS-2022 project TIMES, co-funded
by the European Union. Views and opinions expressed are
however those of the author(s) only and do not necessarily
reflect those of the European Union.}\vspace{-1cm}}

\begin{document}

\maketitle

\begin{abstract}
    With global IMT traffic expected to grow 10-100 times from 2020 to 2030\footnote{Forecasts are reported in \cite{IMT2030}.}, the \textcolor{black}{Terahertz (THz)} spectrum offers a promising solution to satisfy such forecasts. However, occupying the THz spectrum comes with its own challenges, an important one being impairments caused by \textcolor{black}{broadband} RF components in THz transceivers. Nonlinearities in power amplifiers (PAs) complicate meeting link budget requirements, with amplitude and phase distortions degrading the system's performance, especially when adopting waveforms with high peak-to-average power ratios (PAPRs), such as \gls{ofdm}. In this paper, we present characterization results of a 300\,GHz PA using small-signal and large-signal continuous-wave measurements. Models capturing \gls{amam} and \gls{ampm} behavior across 270-330\,GHz are developed and verified with wideband measurements, confirming the compression behavior, while nonetheless showing inaccuracies for low input powers due to unaccounted frequency dependencies. Based on the derived models, a predistortion algorithm is designed and analyzed, revealing significant error performance degradation when switching between single- and multi-carrier waveforms. We finally show that an appropriate selection of predistorter parameters can significantly improve the performance.
\end{abstract}

\begin{IEEEkeywords}
modeling, power amplifiers, pre-distortion, terahertz.
\end{IEEEkeywords}

\vspace{-0.5cm}\section{Introduction}
% The sub-THz frequency range is currently being considered as a means to satisfy the demands on higher data rates. Such high data rates can be achieved thanks to the availability of wide ranges of unused frequency bands. In particular, the frequency range of 250 to 325,GHz is a part of the standards for future wireless communication use cases such as wireless mobile backhauling \cite{IEEE802.15.3d-2023,ITU_WRC2023}. Moreover, several other sub-THz-related use cases have been identified as part of the ETSI ISG THz pre-standardization activity such as interactive immersive XR and the commissioning of industrial plants \cite{10872849,ISGTHzGR002}. To be able to exploit this frequency range, a number of challenges need to be addressed, and one prominent challenge is the radio-frequency (RF) hardware impairments. The same ETSI ISG THz group has recently published a report detailing some of these impairments and specifying models to characterize RF impairments such as oscillator phase noise, antenna beamsquinting and power amplifier non-linearities \cite{ISGTHzGR004}. Such impairments can hinder the full exploitation of the operation at the sub-THz frequency band due to the reduction of the throughput imposed by the untreated transceiver RF hardware impairments.
\gls{thz} frequency bands are currently considered as a means to satisfy the always increasing demand for higher data rates and overcome the spectrum scarcity problem faced by traditional wireless systems~\cite{9681870}. The \gls{itu} already identified a total of 137~GHz between 275 and 450~GHz for mobile and fixed services, and further concessions are expected in the near future~\cite{ITU_WRC2019}. 
The wide bandwidth available at these frequencies can be exploited to deliver extremely high communication performance, achieving data rates up to terabits-per seconds and ultra-low latency. Thanks to the enhanced directionality, \gls{thz} systems can reduce interference and achieve higher spatial reuse. In addition, the small size of antennas at these frequencies enables the fabrication of compact transceivers suitable for small-scale applications. Further, the short wavelength and unique propagation properties of \gls{thz} signals make them ideal for precise sensing and localization, positioning this technology as a promising candidate for implementing integrated sensing and communication functionalities.

The sub-\gls{thz} technology has already been considered for establishing point-to-point and short range wireless connections. The IEEE developed a standard operating between \textcolor{black}{252} and \textcolor{black}{325}~GHz and supporting channel bandwidths up to 69~GHz~\cite{IEEE802.15.3d-2023}. The target applications include wireless front/backhauling, wireless data centers, kiosk downloading, and intra-device communications.
Other potential use cases have been identified by different research and industrial initiatives, including cooperative mobile robots, \gls{xr}, wireless data centers, real-time industrial control at the edge, and virtual commissioning of industrial plants with hardware in the loop~\cite{10872849, 10453153, 9681870}. 

%To be able to exploit this frequency range, a number of challenges need to be addressed, and one prominent challenge is the \gls{rf} hardware impairments. Recently, \gls{etsi} has organized an industry specification group on the topic of \gls{thz} (\gls{etsi} ISG THz), that has recently published a report detailing some of these impairments and specifying models to characterize RF impairments such as oscillator phase noise, antenna beamsquinting and power amplifier non-linearities \cite{ISGTHzGR004}. Such impairments can hinder the full exploitation of the operation at the sub-THz frequency band due to the reduction of the throughput imposed by the untreated transceiver \gls{rf} hardware impairments.

The exploitation of this frequency range comes with a set of challenges, including harsh and variable propagation conditions experienced over wireless channels and pronounced hardware impairments faced by sub-\gls{thz} components. 
In \cite{ISGTHzGR004}, several impairments have been analyzed, including oscillator phase noise, antenna beamsquinting, and power amplifier non-linearities. If not properly treated, such defects can limit the achievable performance and hinder the full exploitation of the sub-\gls{thz} frequency bands. 
In this regard, the \gls{pa} is one of the most critical components in the \gls{rf} chain. Ideally, the \gls{pa} should increase the power of the modulated signal. \textcolor{black}{However, \glspl{pa} operating at sub-\gls{thz} frequencies exhibit pronounced non-linearity effects,} leading to unwanted signal distortion when operating at high power levels. 
On the other hand it is advantageous to operate the \gls{pa} near the compression region due to increased power efficiency.
This is particularly relevant when adopting multi-carrier waveforms such as \gls{ofdm}, in which large amplitude fluctuations in the modulated signal -- high \gls{papr} -- may strain the \gls{pa} and cause significant distortion. Moreover, when operating over wide bandwidths, \glspl{pa} may exhibit memory effects which further distort the input signal.

These effects can be compensated by adopting linearization techniques such as signal predistortion, which applies an inverse transformation to the input signal in the baseband domain before entering the power amplification stage. The first step for implementing this technique is to evaluate the non-linear behavior of the \gls{pa} and develop suitable amplitude and phase distortion models. A common approach is to conduct measurement for characterizing the \gls{pa} device and use the empirical data to extract suitable behavioral models describing the input-output relationships. 

\subsection{Pre-existing models and prior art}
For the evaluation of the NR waveform performance, the 3GPP adopted the polynomial model described in \cite{3gpp20205g}. Although it has been derived from empirical measurements at sub-6 GHz, this model has been recommended by 3GPP also for carrier frequencies above 6 GHz. Since it has been derived from measurements with input power between -35 and 9\,dBm, it may not provide accurate results outside this range.
An alternative Rapp model for above-6 GHz evaluations was proposed in \cite{3gpp20205g2}, however, it was considered less appropriate than the sub-6 GHz polynomial model since, although being more accurate than a simple clipping model, the applicability of the Rapp approach was questioned in the case of wideband operations.

Metrology and device characterization in the sub-\gls{thz} band is restricted mostly to \gls{cw} measurements or two-tone measurements. \textcolor{black}{At high millimeter-wave frequencies, measurements} are performed mostly indirectly with \textcolor{black}{frequency} extension modules or other frequency conversion units, while analyzing the data at much lower frequency.
In \cite{HEXAXIID23,singh2021290}, authors measured the input-output relationship of a SiGe \gls{pa} operating at 290\,GHz. The results were used to derive the parameters of popular memoryless/quasi-memoryless models, including Rapp, Ghorbani, White, and Saleh models. The authors claim that the Rapp model provides the best trade-off between accuracy and model validity, as it shows a meaningful behavior for input power levels outside of the measurement range. 
%In this paper, using various calibration methods, accurate measurements are enabled. Therefore, such a device characterization approach enables accurate modeling, verification, and drawing conclusions on the limits imposed by \glspl{pa} on the performance of \gls{thz} communication systems. 

 %To verify the derived models with actual wideband measurements, we make use of a novel cross-domain characterization method, explained in more detail in chapter\,\ref{ch:characterization}. This setup is able to reduce the influence of the measurement setup and the thermal noise contribution to a negligible level, making an isolated view on the \gls{dut} possible while using wide-band modulated data. This enables a range of possibilities, like more detailed system-simulations or even considerations regarding analog or digital predistortion of the \gls{pa}, as presented in chapter\,\ref{sec: PD}.
%Hence, system-level device modeling (i.e. reported in \cite{HEXAXIID23,singh2021290}) is based on the AM-AM and AM-PM characteristics from the \gls{cw} data. However, considering broadband communication, it is necessary to take into account the wideband effects of \gls{dut}.

%Additionally, a method for simulating the input-output relationship of \glspl{pa} was introduced. This solution can be used for building PA models accounting for memory and non-linear effects. The method has been validated using measurements of a GaN \gls{pa}.

Recently, the \gls{etsi} has established an industry specification group on the topic of \gls{thz}, named \gls{etsi} ISG THz, where in a recently published report on \gls{rf} impairments \cite{ISGTHzGR004}, it extended the modified Rapp model proposed in \cite{3gpp20205g2} by collecting the performance metrics of several sub-\gls{thz} \glspl{pa} operating between 100 and 200~GHz that are available in the literature. The collected data was used to derive the parameters of the average AM-AM distortion characteristic.

The non-linear \gls{pa} behavior is considered as a key issue for enabling efficient and reliable \gls{thz}/sub-\gls{thz} communications~\cite{tarboush2022single, lee2020sub}. Previous works compared the performance of different \gls{phy} layer waveforms under \gls{pa} non-idealities. The Rapp model proposed by 3GPP was used in~\cite{HE20241297} and \cite{maki2023generation} for the evaluation of different single and multi-carrier waveforms. 
In~\cite{tubbax2004ofdm}, authors used a simple cubic \gls{pa} model to compare the performance of \gls{ofdm} and single carrier waveforms in a multi-antenna system. 
%The results show that the impact of the non-ideal \gls{pa} behavior is negligible compared to the frequency-selective effect caused by multi-path channels. 
The authors of \cite{parisi2023modulations} presented an analysis of different modulation schemes for \gls{thz} communications where the saturation effect of the \gls{pa} is considered by applying a \gls{papr} penalty to the input signal. Results show that low-order modulation schemes perform best but pay in terms of spectral efficiency.

\subsection{\textcolor{black}{Contributions of this work}}
 %To verify the derived models with actual wideband measurements, we make use of a novel cross-domain characterization method, explained in more detail in chapter\,\ref{ch:characterization}. This setup is able to reduce the influence of the measurement setup and the thermal noise contribution to a negligible level, making an isolated view on the \gls{dut} possible while using wide-band modulated data. This enables a range of possibilities, like more detailed system-simulations or even considerations regarding analog or digital predistortion of the \gls{pa}, as presented in chapter\,\ref{sec: PD}.
%Hence, system-level device modeling (i.e. reported in \cite{HEXAXIID23,singh2021290}) is based on the AM-AM and AM-PM characteristics from the \gls{cw} data. However, considering broadband communication, it is necessary to take into account the wideband effects of \gls{dut}.
Characterizing \glspl{pa} operating at sub-\gls{thz} frequencies presents significant challenges due to difficulties in the measurement process. Measurement benches capable of properly isolating the \gls{pa} behavior and providing accurate results at such high frequencies are difficult to set up. 
The lack of measurement data prevents the extrapolation of suitable behavioral models and their validation. Most of the models available in the literature are derived from measurements taken at lower frequencies which are unable to properly capture the \gls{pa} behavior at sub-\gls{thz}. 
%The absence of accurate and validated models hampers the development of effective compensation strategies, such as predistortion techniques, which are crucial for mitigating non-linearities and improving the performance of \gls{thz} systems.

In this paper, a characterization of a \gls{pa} operating at 300~GHz is performed based on small-signal and large-signal continuous-wave measurements. \textcolor{black}{The tested device is a 300-GHz Indium Gallium Arsenide (InGaAs) metamorphic high-electron mobility transistor (mHEMT) \gls{pa} \cite{John2019_IMS,John2020_TThz_300G_PA_MMICs}}. Several models are derived that yield AM-AM and AM-PM behavior for several frequency bands spanning the 270 to 330\,GHz frequency range. Measurements obtained by adopting a \gls{vna}-based cross-domain measurement platform \textcolor{black}{(Reported in \cite{Kallfass_IMS2023_CrossLink}, \cite{Schoch_EuMW2023_CrossLink_WBand}, \cite{Teyssier_2023_VectorandStiching} and \cite{ILHLit:NLMdl_RWW_Paper})} demonstrate novel characterization capabilities under broadband-modulated conditions and compare the simulation results based on the derived models with the measurements. Based on the derived models, a predistortion algorithm is derived and tested considering single and multi-carrier waveforms while varying various performance metrics.

The contributions presented in this paper are summarized as follows:
\begin{itemize}
    \item We make use of a novel cross-domain characterization method, \textcolor{black}{which} is able to reduce the influence of the measurement setup and the thermal noise contribution to a negligible level, making an isolated view on the \gls{dut} possible while using wideband modulated data. In Section\,\ref{ch:characterization}, we present detailed measurement data showing significant changes in the behavior of a \gls{thz}-\gls{pa} over the frequency of operation. This data is publicly available \cite{Model_data} to facilitate possible standardization processes.
    \item  In Section\,\ref{chapter_Modelling}, we propose multiple system-level models and evaluate them comprehensively on the basis of the measured data. We provide the model parameters and make them publicly available in \cite{Model_data}, and verify the derived models using results from wideband measurements. We further perform a comparison with different available models in the literature and discuss this comparison. 
    \item In Section \ref{sec: PD}, a predistortion algorithm is designed based on the derived models. Implications on the application of predistortion on the performance of single- and multi-carrier waveforms are shown, thus addressing an open question on the applicability of multi-carrier waveforms at such frequencies. 
\end{itemize}
\textcolor{black}{Initial results of this work were presented in \cite{ILHLit:NLMdl_RWW_Paper}, where the same measurement setup was used to characterize the 300-GHz InGaAs mHEMT power amplifier at a single operating frequency (280 GHz), and only the parameters for the polynomial and the modified Rapp models were derived. On the other hand, in this study we report new results for several frequencies within the 60 GHz bandwidth from 270 to 330 GHz, and we use these measurements to derive the parameters of the polynomial and modified Rapp models, along with the Saleh and Ghorbani models, which were not taken into consideration in \cite{ILHLit:NLMdl_RWW_Paper}. Also, the analysis of communication systems employing pre-distortion algorithms to compensate the PA non-linearities was not present in \cite{ILHLit:NLMdl_RWW_Paper}.}

% \begin{itemize}
%     \item characterization of THz PA using novel measurement methodology (Simon?)
%     \item modelling of PA behavior AM-AM AM-PM characteristics over different operating frequencies. Database publicly available
%     \item validation of the models using simulations and real measurements with wideband signals
%     \item performance evaluation for SC and MC waveforms based on the derived models with and without PA compensation using PD
% \end{itemize}

%\subsection{Paper Organization}

\section{Communication System Model}

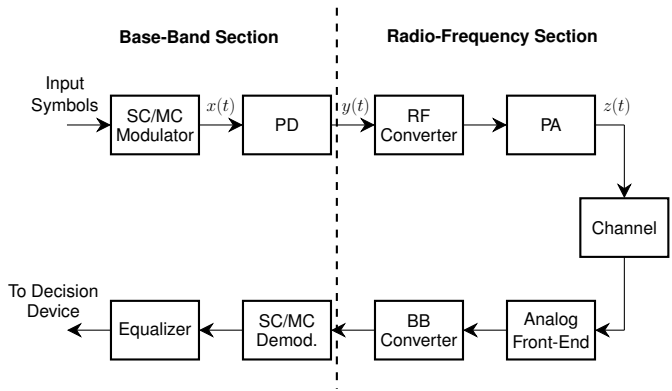
\begin{figure}[t!]
\begin{center}
\begin{tikzpicture}[scale=0.39, every node/.style={scale=0.39}]
[every text node part/.style={align=center}]

\node[left][align = center] at (1.7,1.0) {\LARGE \textsf{Input}\\[6 pt] \LARGE  \textsf{Symbols}};
\draw[{}-{Stealth[length=2mm,width=2mm]}] (0.5,0) -- (2,0);

\draw[thick]  (2,-1) rectangle (5,1);
\node[align = center] at (3.5,0) {\LARGE \textsf{SC/MC}\\[6 pt] \LARGE  \textsf{Modulator}};

\draw[{}-{Stealth[length=2mm,width=2mm]}] (5,0) -- (6.5,0);
\node[above] at (5.75,0.2) {\LARGE $x(t)$};

\draw[thick]  (6.5,-1) rectangle (9.5,1);
\node[align = center] at (8.0,0) {\LARGE  \textsf{PD}};

\draw[{}-{Stealth[length=2mm,width=2mm]}] (9.5,0) -- (11,0);
\node[above] at (10.35,0.2) {\LARGE $y(t)$};

\draw[thick]  (11,-1) rectangle (14,1);
\node[align = center] at (12.5,0) {\LARGE \textsf{RF}\\[6 pt] \LARGE \textsf{Converter}};

\draw[{}-{Stealth[length=2mm,width=2mm]}] (14,0) -- (15.5,0);
% \node[above] at (14.75,1.0) {\LARGE $y_{\rm RF}(t)$};

\draw[thick]  (15.5,-1) rectangle (18.5,1);
\node[align = center] at (17,0) {\LARGE \textsf{\gls{pa}}};

\node[above] at (19.25,0.2) {\LARGE $z(t)$};

\draw[{}-{}] (18.5,0) -- (19.5,0);
\draw[{}-{Stealth[length=2mm,width=2mm]}] (19.5,0) -- (19.5,-2.5);

\draw[thick]  (18,-2.5) rectangle (21,-4.5);
\node[align = center] at (19.5,-3.5) {\LARGE \textsf{Channel}};

\draw[{}-{}] (19.5,-4.5) -- (19.5,-7);
\draw[{}-{Stealth[length=2mm,width=2mm]}] (19.5,-7) -- (18.5,-7);

\draw[thick]  (15.5,-8) rectangle (18.5,-6);
\node[align = center] at (17,-7) {\LARGE \textsf{Analog} \\[6 pt] \LARGE \textsf{Front-End}};

\draw[{}-{Stealth[length=2mm,width=2mm]}] (15.5,-7) -- (14,-7);

\draw[thick]  (11,-8) rectangle (14,-6);
\node[align = center] at (12.5,-7) {\LARGE \textsf{BB}\\[6 pt] \LARGE \textsf{Converter}};

\draw[{}-{Stealth[length=2mm,width=2mm]}] (11,-7) -- (9.5,-7);

\draw[thick]  (6.5,-8) rectangle (9.5,-6);
\node[align = center] at (8,-7) {\LARGE \textsf{SC/MC}\\[6 pt] \LARGE  \textsf{Demod.}};

\draw[{}-{Stealth[length=2mm,width=2mm]}] (6.5,-7) -- (5,-7);

\draw[thick]  (2.0,-8) rectangle (5,-6);
\node[align = center] at (3.5,-7) {\LARGE \textsf{Equalizer}};

\draw[{}-{Stealth[length=2mm,width=2mm]}] (2,-7) -- (0.5,-7);

\node[left][align = center] at (1.7,-6.0) {\LARGE \textsf{To Decision}\\[6 pt] \LARGE  \textsf{Device}};

\draw[thick,dashed] (9.7,4) -- (9.7,-9);
\node at (5,3) {\LARGE \textbf{\textsf{Base-Band Section}}};
\node at (15,3) {\LARGE \textbf{\textsf{Radio-Frequency Section}}};

\end{tikzpicture}
\end{center}
\caption{Schematic diagram of the communication system under investigation.}
\label{fig: SystemModel}
\end{figure}

Fig.~\ref{fig: SystemModel} shows a schematic diagram of the communication system under investigation. The input symbols, which typically belong to an $M$-QAM constellation, are fed to a modulator that produces the continuous-time waveform, $x(t)$. Here we consider both single-carrier and multi-carrier modulations.  The latter are efficiently implemented through an \gls{ofdm} architecture which uses fast Fourier transform (FFT) algorithms. A detailed description of standard single-carrier and \gls{ofdm} transmitters, with bi-dimensional constellations, can be found in many textbooks, e.g., \cite{Proakis2007}. Essential  components of the communication system in Fig.~\ref{fig: SystemModel} are the predistortion device and the \gls{pa}, which operates at sub-\gls{thz} frequencies. 
% The operation of the \textit{analog} PD, which is used to compensate for the non-linear effects introduced by the \gls{pa}, will be explained and analyzed in Section~\ref{sec: PD}. 
% Measurements results and theoretical models for \gls{pa} will be given and discussed in Section~\ref{ch:characterization} and \ref{chapter_Modelling}, respectively. 
%The predistortion algorithm works on the baseband signal while amplification is performed on the radio-frequency (RF) signal, before transmission over the channel through the transmit antenna.

% The predistortion algorithm are considering the baseband-equivalent of the \gls{rf} signal, focusing on the signal before the transmit antenna, before transmission over the channel.

The predistortion algorithm works on the baseband equivalent of the \gls{rf} signal while amplification is performed on the up-converted signal, before transmission over the channel.

The electromagnetic waveform at the output of the propagation medium is converted to an electric signal by the receive antenna, and then processed through a standard receiver, as shown in the lower part of Fig.~\ref{fig: SystemModel}. In particular, after baseband (BB) conversion and demodulation, the signal samples are equalized before they are sent to the decision device.
  
In considering the communication system reported in Fig.~\ref{fig: SystemModel}, the main objective is that of comparing the performance of single-carrier and multi-carrier waveforms at sub-\gls{thz} frequencies, in the presence of a predistortion algorithm. It is well-known that, with respect to a single-carrier system, the \gls{ofdm} signal is characterized by a much higher \gls{papr}, and this results in severe distortions in passing through the \gls{pa}. In a frequency-flat channel single-carrier systems are undoubtedly the better choice. On the other hand, multi-carrier modulations could be preferable in the presence of a frequency-selective behavior of the propagation medium, due, for example, to frequency-dependent molecular absorption, or simply to an environment with rich scattering, as demonstrated in \cite{10464485}. Accordingly, in the context of sub-THz communications, it becomes very important to accurately analyze the performance of single-carrier and multi-carrier waveforms that consider effective techniques for the compensation of \gls{pa} non-linearities.

%\textcolor{blue}{Delete this, because chapter structure already had been introduced: In sections \ref{ch:characterization} and \ref{chapter_Modelling}, we focus on the characterization and the behavioral modeling of the \gls{pa}, while Section \ref{sec: PD} presents the impact of the \glspl{pa} on waveforms operating at sub-\gls{thz} while also applying predistortion.}

%\textcolor{black}{Simon: Following chapters \ref{chapter_Modelling} and \ref{ch:characterization} are focusing exclusively on the \gls{pa}-device, while chapter \ref{sec: PD} is considering the overall system.}

\section{Characterization of a THz-Power Amplifier}
\label{ch:characterization}

 \begin{figure}[!t]
 	\centering
 	\includegraphics[width=0.9\columnwidth]{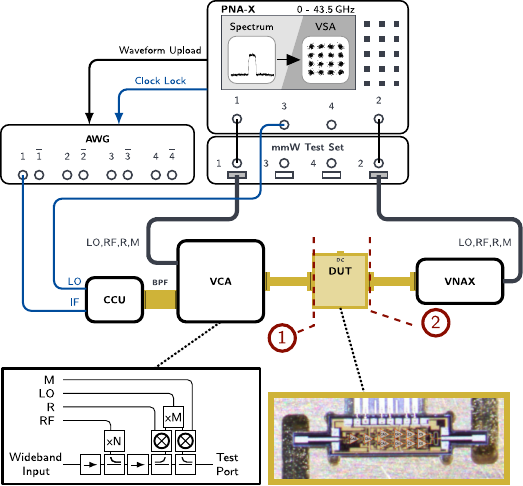}
 	\caption{Block diagram of the cross-domain setup for \gls{cw} as well as wideband-modulated characterization of the H-band amplifier \gls{dut}, with calibration planes 1 at the input and 2 at the output (from \cite{ILHLit:NLMdl_RWW_Paper}).}
 	\label{fig_MeasSetup}
  \vspace{-0.5cm}
 \end{figure}

The scope of this section is focused on the evaluation of the influence of the non-linearities of the \gls{thz}-\gls{pa} for the extraction of system-level models. For device characterization in the sub-\gls{thz} domain, \textcolor{black}{frequency-}extension modules are usually used to increase the frequency range of the base unit. Using vector calibration methods, the influence of the up- and down-conversion of the signal as well as other impairments in the signal generation or acquisition can be minimized and an isolated view on the \gls{dut} is possible. This procedure can be applied to \gls{cw} characterization of sub-\gls{thz} devices. However, an important aspect of our work is also the measured validation of the extracted system-models, while being excited with wideband modulated data. This is enabled by a cross-domain measurement testbench, which is described in more detail in \cite{Kallfass_IMS2023_CrossLink}, \cite{Schoch_EuMW2023_CrossLink_WBand}, \cite{Teyssier_2023_VectorandStiching} and \cite{ILHLit:NLMdl_RWW_Paper}. Fig.\,\ref{fig_MeasSetup} shows the block diagram of the characterization setup.
The input and the output of the \gls{dut} are connected via WR-3.4 waveguide interfaces to the \textcolor{black}{VNAX (\gls{vna} Extension Module) and VCA (Vector Component Analyzer). The VNAX  is a standard waveguide extension, whereas the VCA} has an additional interface for injecting broadband signals towards the \gls{dut}. The extenders are connected to port\,1 and port\,2 of a \gls{vna} with spectrum-analyzer option, enabling \gls{cw} characterization as well as broadband signal acquisition in frequency domain. A wideband modulated signal, replayed in a continuous loop by an \gls{awg} is up-converted to a center frequency within the WR-3.4 waveguide band. Usage of an local oscillator from port\,3 of the \gls{vna} and a sensitive clock synchronization of the \gls{awg} are a premise for phase consistency throughout the setup. A band-pass filter filters unwanted \gls{lo} leakage or mirror images from the up-conversion. The amplitudes and phases of the wideband signal at port\,1 and port\,2 are measured. Applying an Inverse Fourier Transform, the time domain representation of the signals can be composed. Using the receiver at port\,1, at the input of the \gls{dut}, the test signal can be corrected to cancel the impairments of the upconversion stage. With this approach, an isolated view on the \gls{dut} with wideband signals is enabled in almost distortion-free conditions.

\subsection{Device Under Test: THz Power Amplifier}
The \gls{dut} used in this work is a 300~GHz WR-3.4 \gls{pa} module similar to \cite{John2019_IMS,John2020_TThz_300G_PA_MMICs}. The \gls{pa} circuit is realized in a 6-stage topology with cascode and common-source devices , utilizing a 35-nm InGaAs metamorphic high electron mobility transistor (mHEMT) technology. The \gls{pa} module is implemented in a split-block packaging technology. An E-field-probe-based transition from the \gls{mmic} to the WR-3.4 waveguide, is implemented directly on the GaAs substrate.  The \gls{pa} reaches on average 20\,dB gain and 7\,dBm of saturated output power within its band of operation from 270\,GHz to over 330\,GHz.

\begin{figure}[]
	\centering
	\subfloat[]{\includegraphics[width=0.98\columnwidth]{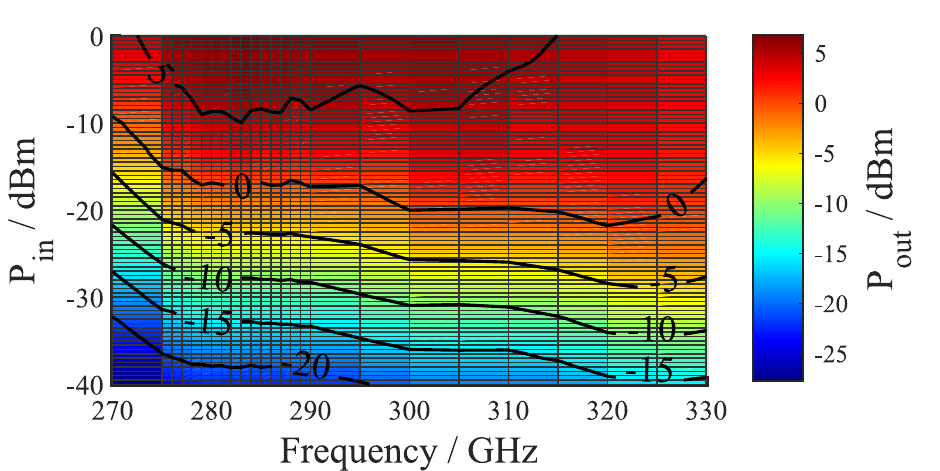}
		\label{fig_PA_Meas_CW_Gain}}
	%\hfil
    \\~\\
	\subfloat[]{\includegraphics[width=0.98\columnwidth]{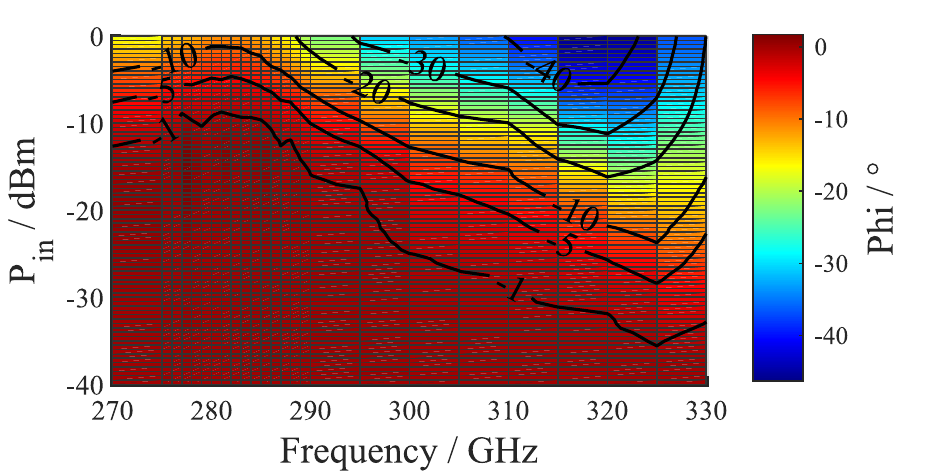}
		\label{fig_PA_Meas_CW_Phase}}
	\caption{Measured \gls{cw} large signal behavior showing (a) gain compression over input power and frequency and (b) phase change of the output tone related to input power and center frequency.}
	\label{fig_PA_Meas_CW}
\end{figure}

% \begin{figure}[]
% 	\centering
%     \includegraphics[width=0.98\columnwidth]{Figures/Characterization/Measurement_Results_Constellations.pdf}
%     \caption{Measured constellation diagrams for QPSK, 16-QAM and 64-QAM at the input and output of the \gls{dut} are given for 1\,GBd symbol rate. The reference planes 1 and 2 are corresponding to the planes in Fig.\,\ref{fig_MeasSetup}}
% 	\label{fig:PA_Meas_wideband}
% \end{figure}

{
\begin{table}[]
\caption{Measured constellation diagrams from cross-domain characterization at 280\,GHz center frequency and 1\,GBd symbol rate.}
\begin{tabular}{|p{1.2cm}|p{1.9cm}|p{1.9cm}|p{1.9cm}|}
\hline
     & \textbf{QPSK} & \textbf{16-QAM} & \textbf{64-QAM} \\ \hline
\raisebox{-0\totalheight}{\includegraphics[width=0.6cm]{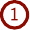}} \vspace{0.2cm} \newline 
Corrected Signal at \gls{pa} input 
& \raisebox{-0.4\totalheight}{\includegraphics[width=1.8cm]{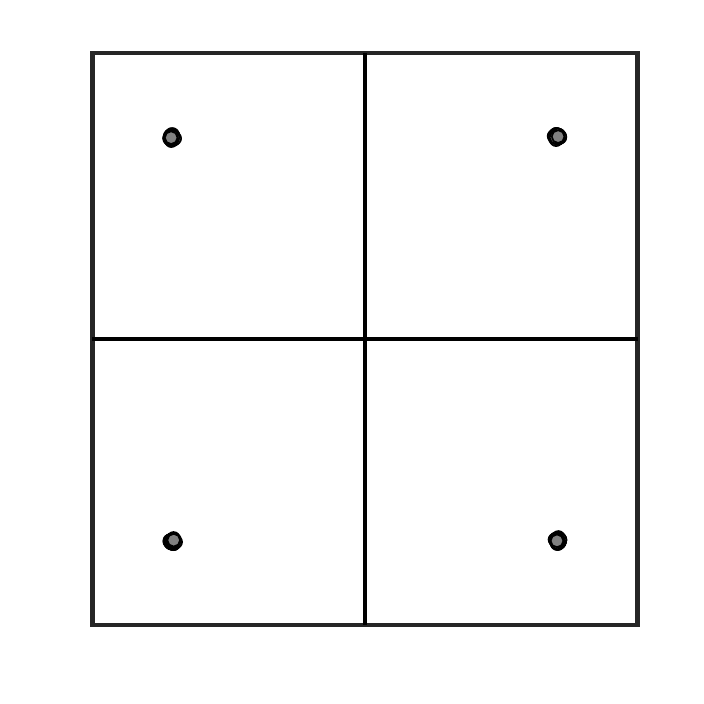}} \newline EVM\,=\,-44.4\,dB \newline $P_{\mathrm{in}}$\,=\,-12.9\,dBm
& \raisebox{-0.4\totalheight}{\includegraphics[width=1.8cm]{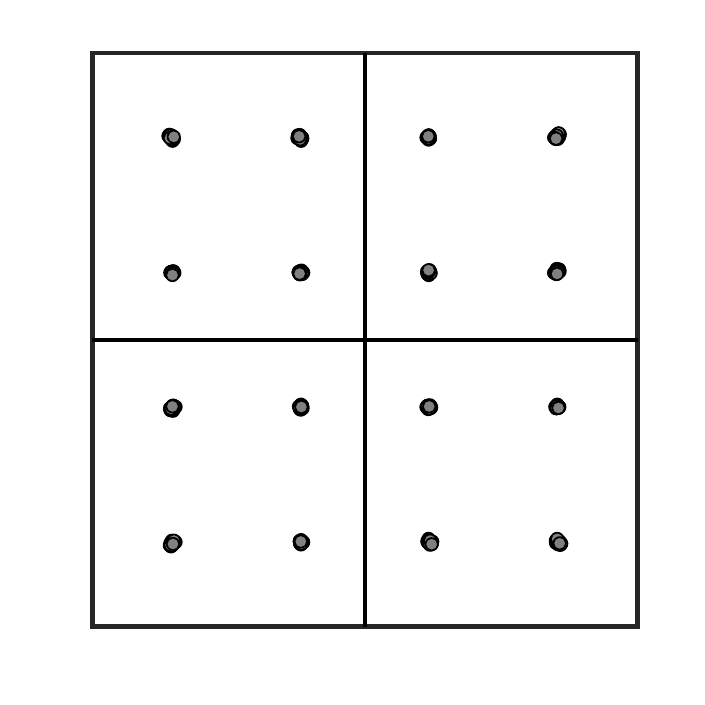}} \newline EVM\,=\,-46.2\,dB \newline $P_{\mathrm{in}}$\,=\,-13.2\,dBm
& \raisebox{-0.4\totalheight}{\includegraphics[width=1.8cm]{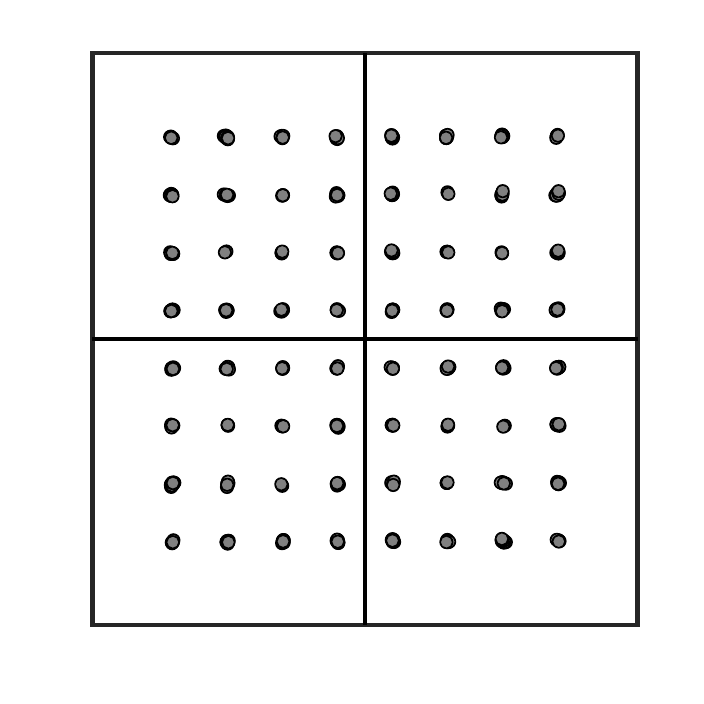}}\newline EVM\,=\,-43.5\,dB \newline $P_{\mathrm{in}}$\,=\,-13.1\,dBm
\\ \hline
\raisebox{-0\totalheight}{\includegraphics[width=0.6cm]{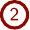}} \vspace{0.2cm} \newline
Signal at \gls{pa} output 
& \raisebox{-0.4\totalheight}{\includegraphics[width=1.8cm]{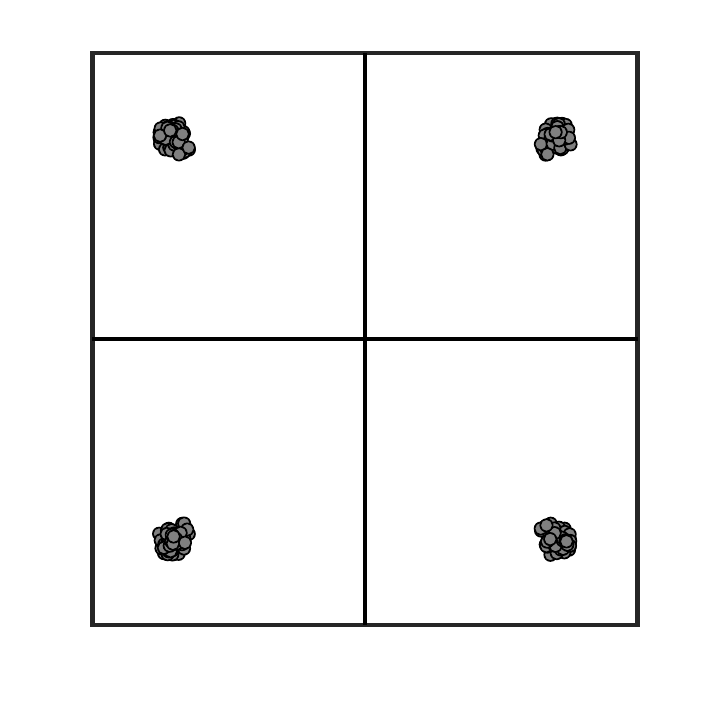}} \newline EVM\,=\,-29.7\,dB \newline $P_{\mathrm{out}}$\,=\,+3.3\,dBm
& \raisebox{-0.4\totalheight}{\includegraphics[width=1.8cm]{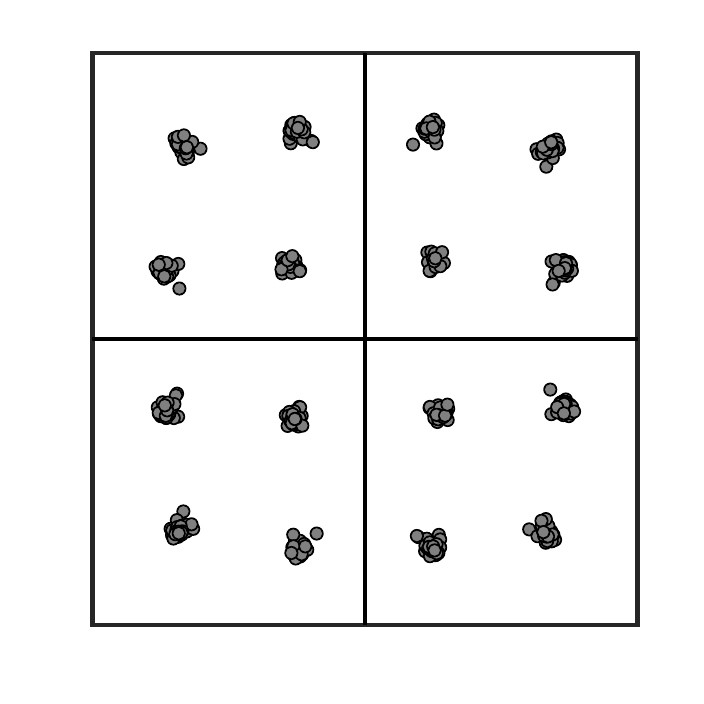}} \newline EVM\,=\,-26.2\,dB \newline $P_{\mathrm{out}}$\,=\,+3.0\,dBm
& \raisebox{-0.4\totalheight}{\includegraphics[width=1.8cm]{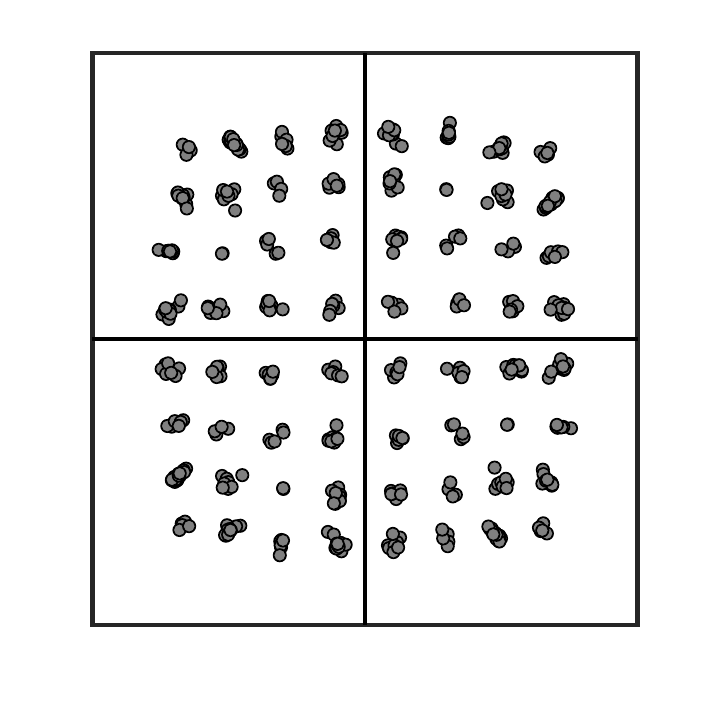}}\newline EVM\,=\,-26.9\,dB \newline $P_{\mathrm{out}}$\,=\,+2.8\,dBm
\\ \hline
\end{tabular}
\label{tab:PA_Meas_wideband}
%\vspace{-\baselineskip}% remove one line of space below this table
\end{table}
}

\subsection{Characterization}
Fig.\,\ref{fig_PA_Meas_CW} shows the vectorial \gls{cw}-characterization of the \gls{pa} over both frequency and input power $P_{\mathrm{in}}$, in terms of AM-AM and AM-PM relations. Sub-figure (a) shows the compression of the output power $P_{\mathrm{out}}$. A saturated output power of approximately 7\,dBm is reached. The curve indicates the varying gain over the frequency and the non-constant spacing between the equi-power lines attests to different compression curves over frequency. Further, in sub-figure (b), the phase modulation is plotted, which is showing major dependency between the phase modulation and the operation frequency.
The normalized phase shift $\Phi$ is defined as the difference of the measured phase shift, compared to the phase shift measured at the linear operation point with $P_{\mathrm{in}} = -40\text{ dBm}$  at the same frequency, which can be written as
\begin{equation}
    \Phi(f,P_{\mathrm{in}}) = \phi_{2,1}(f,P_{\mathrm{in}}) -\phi_{2,1}(f,P_{\mathrm{in}}=-40\text{ dBm}).
\end{equation}
This AM-AM and AM-PM data is used in Section~\,\ref{chapter_Modelling} to extract the behavioral models.\\
However, to assess the modulation quality and to compare the behavioral models with the actual behavior of the \gls{dut}, broadband measurements are carried out. With a baudrate of 1\,GHz and modulation formats of QPSK, 16-QAM and 64-QAM, the signal quality in terms of \gls{evm} is measured in the cross-link testbench. The \gls{evm} is defined as the \gls{rms} of the error vector  divided by the maximum signal vector of the reference constellation in logarithmic scale. Because of the vectorial calibration capabilities, the influence of the measurement setup, the up- and down-conversion can be corrected from the measurement and solely the influence of the \gls{dut} is included. With vector averaging even the noise contribution is reduced to a level, where 45\,dB of signal-to-noise ratio is achieved in the measurement. Some exemplary constellation diagrams, where the \gls{pa} is operated in compression, are shown in the figures in Table \,\ref{tab:PA_Meas_wideband}.

\section{Power Amplifier Non-linearity Modeling}
\label{chapter_Modelling}
 In this section, we derive four different AM-AM and AM-PM distortion models based on empirical narrow-band measurement data using polynomial approximation, Rapp, Ghorbani and Saleh models. 

\subsection{Polynomial amplifier model}
%Similar to the approach adopted in \cite{3gpp20205g}, a polynomial PA model is derived based on AM-AM and AM-PM data provided for a specific operating frequency. The range of root-mean-square (RMS) input powers over which the models are derived is $\mathbb{A}\in[-40,0]$~dBm. It provides output power (RMS) and phase relations to the input power in the dBm domain.

\begin{figure}[t]
    \centering
    \includegraphics[width=0.7\columnwidth]{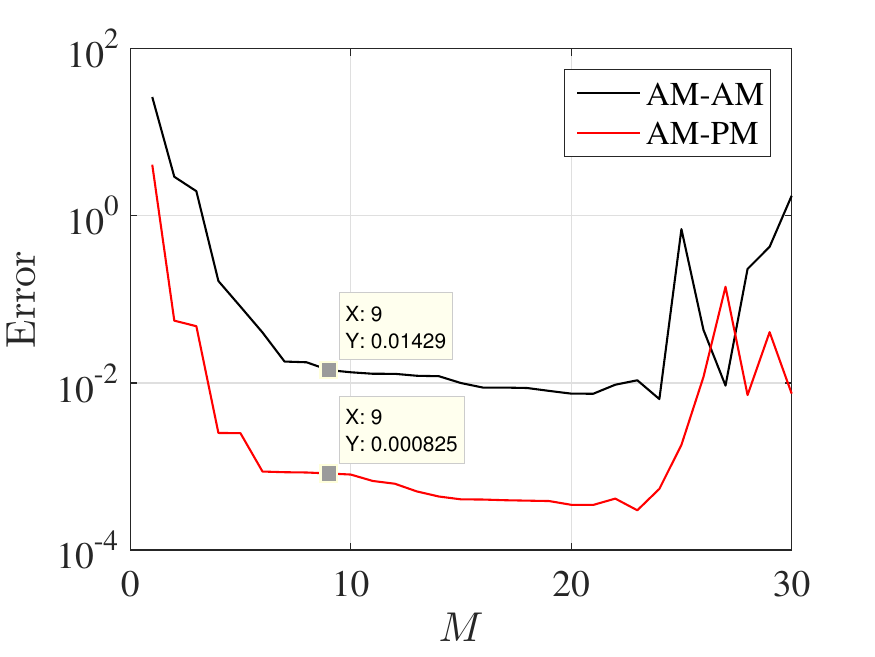}
    \caption{Fitting error of AM-AM and AM-PM polynomial models.}
    \label{fig:poly_err} % \vspace{-0.5cm}
\end{figure}

\begin{figure}[t]
    \centering
    \includegraphics[width=0.8\columnwidth, height=5cm]{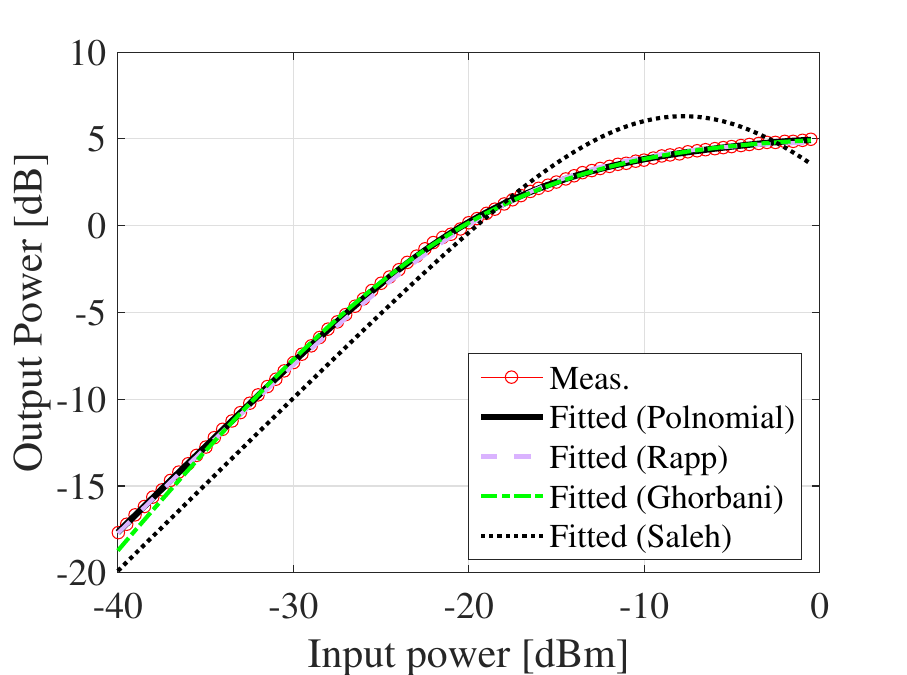}
    \caption{Polynomial, modified Rapp, Ghorbani and Saleh AM-AM models at 315~GHz.}
    \label{fig:All_AMAM} % \vspace{-0.5cm}
\end{figure}

\begin{figure}[t]
    \centering
    \includegraphics[width=0.8\columnwidth, height=5cm]{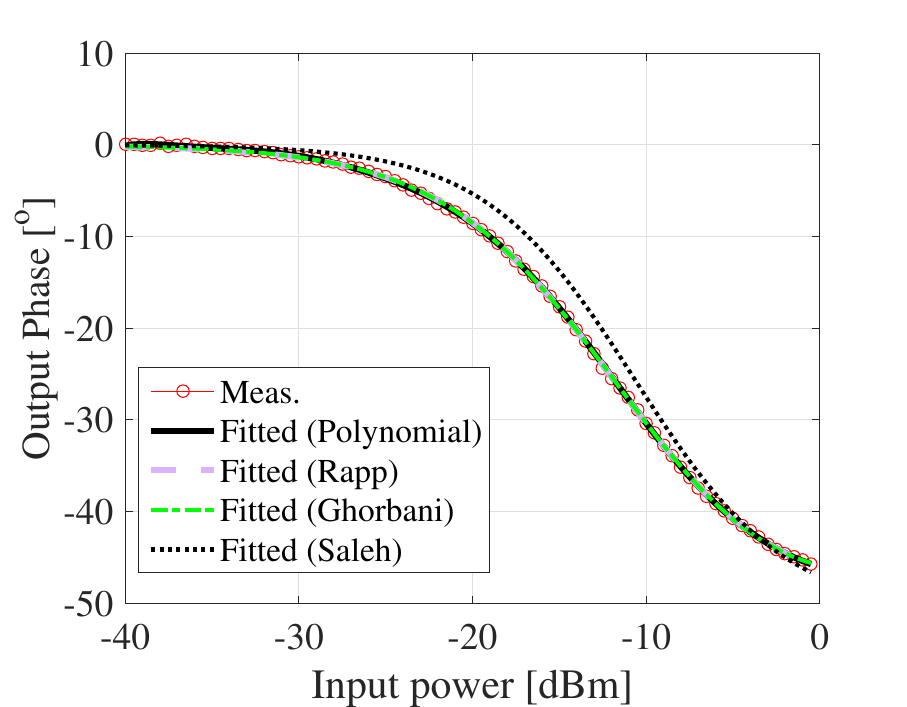}
    \caption{Polynomial, modified Rapp, Ghorbani and Saleh AM-PM models at 315~GHz.}
    \label{fig:All_AMPM} % \vspace{-0.5cm}
\end{figure}

Similar to \cite{3gpp20205g}, a polynomial PA model is derived based on AM-AM and AM-PM data at specific narrow-band operating frequency. The \gls{rms} input power range for this model is $\mathbb{A}\in[-40,0]$~dBm. The model provides the relationship between output power (RMS) and output phase with respect to input power, measured in dBm, which can be written as
\begin{align}
	\alpha_o =\sum_{m=0}^{M} a_m \alpha_i^m, \quad & \quad \vartheta_o =\sum_{m=0}^{M} b_m \alpha_i^m.
\end{align}
\textcolor{black}{The polynomial coefficients are derived based on a least-squares arrangement using the ``polyfit'' function defined in MATLAB$^\copyright$.}
Fig.~\ref{fig:poly_err} shows the error between the model and the measurement data as a function of the polynomial order $M$.  \textcolor{black}{The reason for the poor performance of the polynomial fitting, for polynomial orders above $20$, is the increased sensitivity to rounding errors. Thus, the choice of lower order polynomials should be made both to reduce the complexity and to avoid the loss of accuracy observed with higher order polynomials.} The lowest error is achieved with $M$\,=\,24 (AM-AM) and 23 (AM-PM) however, we consider $M=9$ as in the model provided in \cite{3gpp20205g} which results in an acceptable performance.
% To limit the model complexity, one could reduce the polynomial order $M$ to, e.g., 9 as in the model provided in \tz{add ref} and still obtain acceptable performance. 
\textcolor{black}{Fig.\,\ref{fig:All_AMAM} and \ref{fig:All_AMPM}}
depict the derived models for $f_c$\,=\,315\,GHz and compares them with the measurement. The polynomial coefficients for the polynomial AM-AM and AM-PM behaviors for $f_c=315$ GHz are listed in Table \ref{tab:poly_coeff}. The coefficients relating to the rest of the frequencies are given in \cite{Model_data}.

\begin{table*}
\setlength{\tabcolsep}{5pt}
    \centering
    \caption{Coefficients of polynomial amplifier model.}
% \vspace{-0.3cm}
    \label{tab:poly_coeff}
    \begin{tabular}{c|cccccccccc}
        $m$ & 0 & 1 & 2 & 3 & 4 & 5 & 6 & 7 & 8 & 9\\
        \hline
        %$a_m$ & 6.1583 & 4.3154e-2 & 4.6626e-3 & 2.9347e-3 & 2.0928e-4 & 1.4127e-5 & 6.9011e-7 & 1.8988e-8 & 2.6284e-10 & 1.4227e-12 \\
        %$a_m$ & 6.5703	&-9.65645e-3	&-3.97764e-2	&-7.79282e-3	&-1.23377e-3	&-9.54502e-5	&-4.10498e-6	&-1.01101e-7	&-1.33672e-9	&-7.36348e-12 \\
        $a_m$ & 4.93685	&-0.0137525 &-0.0376565	&-0.00671547	&-0.000751183	&-4.90554e-05	&-2.02848e-06	&-5.08754e-08	&-6.93973e-10	&-3.92275e-12 \\
        %$b_m$ & -23.0789 & -1.5459 & 1.8101e-1 & 2.0984e-2 & 4.03e-4 & -3.0924e-5 & -1.85e-6 & -3.9963e-8 & -3.726e-10 & -1.0996e-12\\
        %$b_m$ & -11.62809	&-1.40685	&-4.45072e-4	&-5.40662e-3	&-2.17524e-3	&-2.09761e-4	&-9.70111e-6	&-2.42573e-7	&-3.16991e-9	&-1.70366e-11 \\
        $b_m$ &-46.00981	&-0.475385	&0.172884 &0.029412	&0.00550807	&0.000508238	&2.42772e-05	&6.33498e-07	&8.63223e-09	&4.82313e-11  \\
    \end{tabular}
% \vspace{-0.5cm}
\end{table*}

% \begin{figure}[]
% 	\centering
% 	\subfloat[]{\includegraphics[width=0.5\columnwidth]{Figures/Models/poly_AMAM_315.eps}\label{poly_AMAM}}
% 	%\hfil
% 	\subfloat[]{\includegraphics[width=0.5\columnwidth]{Figures/Models/poly_AMPM_315.eps}\label{poly_AMPM}}
% 	\caption{Fitted polynomial model (a) AM-AM and (b) AM-PM behavior at 315 GHz.}\label{fig:poly_model}
% \end{figure}

\subsection{Ghorbani model}
The Ghorbani PA nonlinearity model \cite{ghorbani1991effect} characterizes both the AM-AM ($G_A(x)$) and AM-PM ($G_P(x)$) responses using a four parameter equations of the form
\begin{align}
    G_A (x) &=\frac{y_1 x^{y_2}}{( 1 + y_3 x^{y_2} )} + y_4  x,  \\ 
    G_P (x) &=\frac{z_1 x^{z_2}}{( 1 + z_3 x^{z_2} )} + z_4  x, \label{eq:Ghorbani_model}    
\end{align}
where the parameter $y_i, \ z_i, \ i\in \{1,2,3,4\}$, are estimated from the measurements data using curve fitting techniques. The fitting results are depicted in \textcolor{black}{Figs. \ref{fig:All_AMAM} and \ref{fig:All_AMPM}.} 
\textcolor{black}{To derive the fitting parameters, we adopt the second norm of the difference between the raw measurement data and the Ghorbani model as a cost function, i.e.
\begin{align}
    \varepsilon = \norm{ \mathbf{d}_{\mathcal{M}} - \mathbf{g}_{\mathcal{M}} }_2 ,
\end{align}
where $\mathbf{d}$ is the raw measurement data vector, $\mathbf{g}$ is the Ghorbani model vector, $\mathcal{M}\in {A,P}$ with $A$ and $P$ representing the AM-AM and AM-PM functions, respectively. Then, we use the  MATLAB$^\copyright$ core function ``fminsearch'', which is a nonlinear programming solver, to find the fitting parameters of the model.}
The Ghorbani AM-AM and AM-PM coefficients for $f_c$\,=\,$315$\,GHz were found to be: $y_1 = 101.934$, $y_2=1.26$, $y_3=1728.859$, $y_4=-0.0174$, and $z_1 = -1.667\times 10^{5}$, $z_2=1.678$, $z_3=2.981 \times 10^{3}$, $z_4=1.418 \times 10^{2}$. The coefficients for the rest of the operating frequencies are available in \cite{Model_data}. \textcolor{black}{Although the derived parameters properly fit the measurements data as shown in Figs. \ref{fig:All_AMAM} and \ref{fig:All_AMPM}, we point out that the global optimum cannot be guaranteed since the optimization problem is underdetermined. Other optimization tools can be used to derive the parameters.}

% \begin{figure}[]
% 	\centering
% 	\subfloat[]{\includegraphics[width=0.5\columnwidth]{Figures/Models/Ghorbani_AMAM_315.eps}\label{Ghorbani_AMAM}}
% 	%\hfil
% 	\subfloat[]{\includegraphics[width=0.5\columnwidth]{Figures/Models/Ghorbani_AMPM_315.eps}\label{Ghorbani_AMPM}}
% 	\caption{Fitted Ghorbani model (a) AM-AM and (b) AM-PM behavior at 315 GHz.}\label{fig:Ghorbani_model}
% \end{figure}

\subsection{Saleh model}
The Saleh PA nonlinearity modeling \cite{saleh1981frequency} was specifically derived for traveling-wave-tube amplifiers. The AM-AM and AM-PM equations can be implemented by respectively applying
\begin{align}
    S_A (x) &=\frac{\alpha_1 x }{( 1 + \beta_1 x^2 )}, \\ 
    S_P (x) &=\frac{\alpha_2 x^2 }{( 1 + \beta_2 x^2 )}.    
\end{align}
Given their simple forms, Saleh proposed a general form of $S_A(x)$ and $S_P(x)$ given by
\begin{align}
    z(x) = \frac{\alpha x^n}{( 1 + \beta x^2 )^\nu},
\end{align}
where $n\in \{1,2,3\}$, $\nu\in \{1,2\}$. Then, defining $w_m = \frac{1}{\sqrt[\nu]{\frac{z_m}{x_m^n}}}$, the optimal coefficients $\alpha$ and $\beta$ in the mean-square-error sense may be written as \cite{saleh1981frequency}
\begin{align}
 \alpha &=\left[\frac{\left(\Sigma_{m=1}^{N} x_m^2\right)^2-N \Sigma_{m=1}^{N} x_m^4}{\left(\Sigma_{m=1}^{N} x_m^2\right)\left(\Sigma_{m=1}^{N} w_m x_m^2\right)-\left(\Sigma_{m=1}^{N} x_m^4\right)\left(\Sigma_{m=1}^{N} w_m\right)}\right]^\nu \label{eq:alpha} \\
 \beta &=\frac{\left(\Sigma_{m=1}^{N} x_m^2\right)\left(\Sigma_{m=1}^{N} w_m\right)-N \Sigma_{m=1}^{N} w_m x_m^2}{\left(\Sigma_{m=1}^{N} x_m^2\right)\left(\Sigma_{m=1}^{N} w_m x_m^2\right)-\left(\Sigma_{m=1}^{N} x_m^4\right)\left(\Sigma_{m=1}^{N} w_m\right)},\label{eq:beta}
\end{align}
where $N$ denotes the size of the input power vector. Based on the above derivations, the values of the Saleh model parameters at $f_c = 315$ GHz are $\alpha_1=10.127$, $\beta_1=5.995\times 10^{3}$ with $n=1,\ \nu=1$ and $\alpha_2=-595236.026$, $\beta_2 = 11640.052$ with $n=2,\ \nu=1$. The rest of the derived parameters operating at different center frequencies are provided in \cite{Model_data}.

%The values of $\alpha_{\{1,2\}}$, $\beta_{\{1,2\}}$ are shown in Table \ref{tab:Saleh_coeffs} presented in Section \ref{sec:appendix-a5} and were calculated based on (\ref{eq:alpha}) and (\ref{eq:beta}) for different operating frequencies. As an example, the fitted AM-AM and AM-PM behaviors are plotted in Fig. \ref{fig:Saleh_model} for the operating frequency $280$ GHz. 

% \begin{figure}[]
% 	\centering
% 	\subfloat[]{\includegraphics[width=0.5\columnwidth]{Figures/Models/Saleh_AMAM_315.eps}\label{Saleh_AMAM}}
% 	%\hfil
% 	\subfloat[]{\includegraphics[width=0.5\columnwidth]{Figures/Models/Saleh_AMPM_315.eps}\label{Saleh_AMPM}}
% 	\caption{Fitted Saleh model (a) AM-AM and (b) AM-PM behavior at 315 GHz.}\label{fig:Saleh_model}
% \end{figure}

%\begin{figure}[t]
 %   \centering
  %  \includegraphics[width=1\columnwidth]{Figures/Models/Model_comparison.eps}
   % \caption{Comparisons between different AM-AM models.}
    %\label{fig:comparison_AMAM} % \vspace{-0.5cm}
%\end{figure}

% \begin{figure}[]
% 	\centering
% 	\subfloat[]{\includegraphics[width=0.5\columnwidth]{Figures/Models/Rapp_AMAM_315.eps}\label{Rapp_AMAM}}
% 	%\hfil
% 	\subfloat[]{\includegraphics[width=0.5\columnwidth]{Figures/Models/Rapp_AMPM_315.eps}\label{Rapp_AMPM}}
% 	\caption{Fitted Rapp model (a) AM-AM and (b) AM-PM behavior at 315 GHz.}\label{fig:Rapp_model}
% \end{figure}

\begin{figure}[]
	\centering
	\subfloat[]{\includegraphics[width=0.45\columnwidth, height=3.3cm]{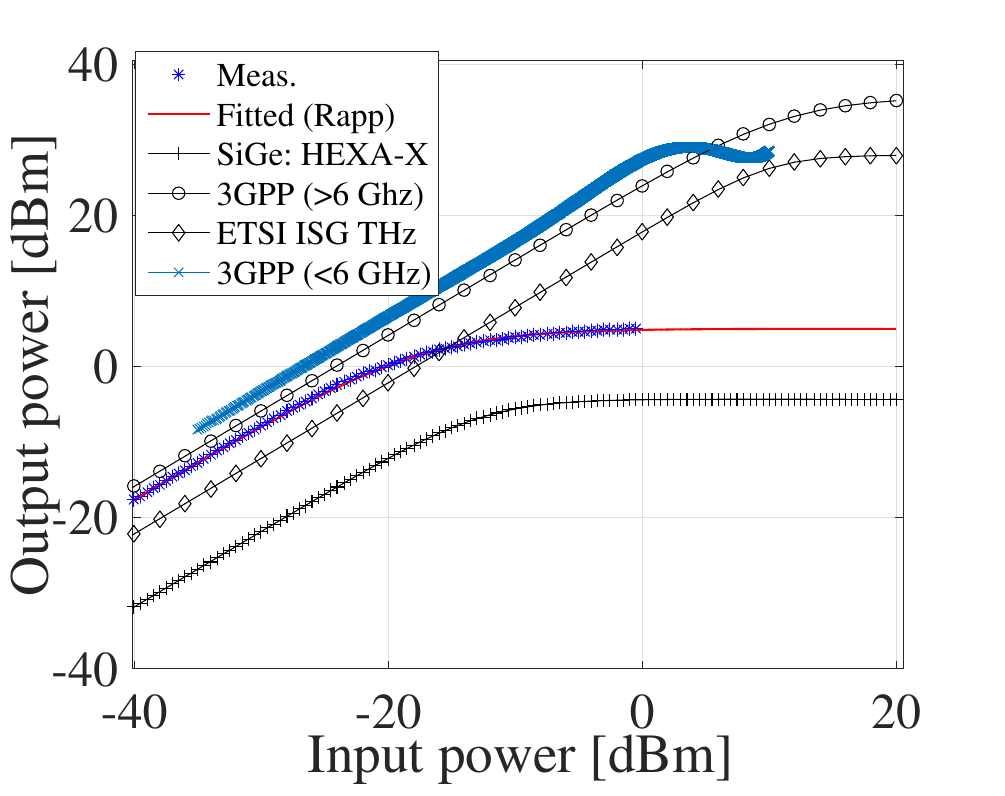}}\label{comparison_AMAM}
	%\hfil
	\subfloat[]{\includegraphics[width=0.45\columnwidth,height=3.3cm]{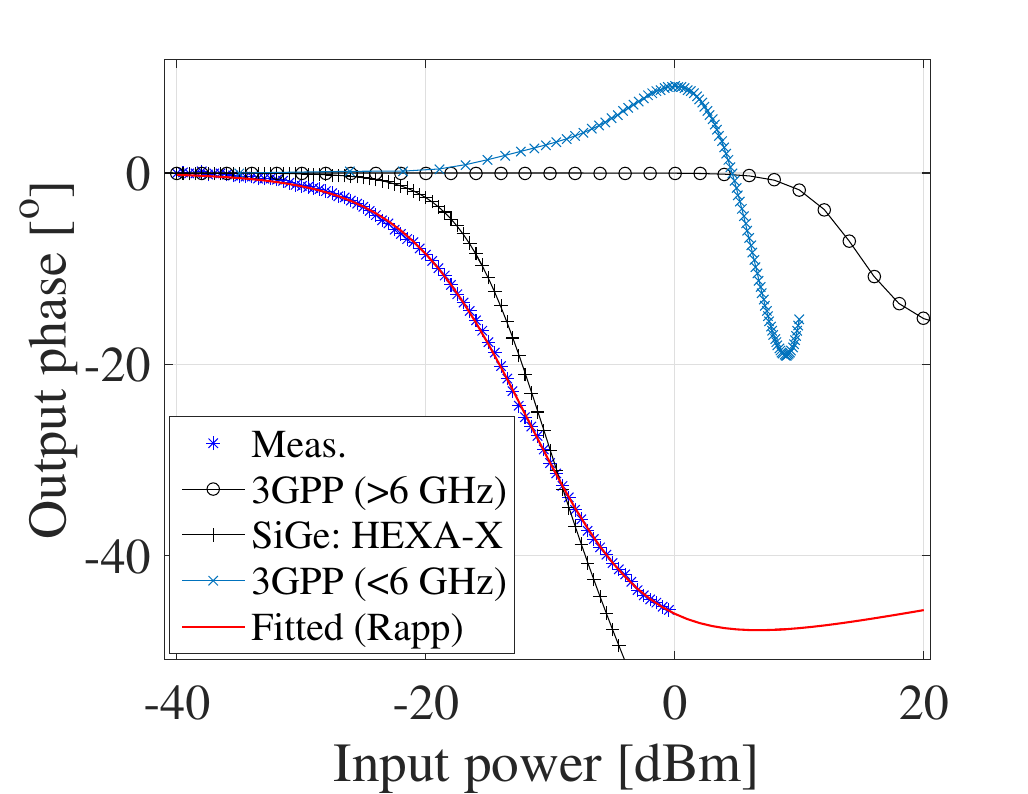}\label{comparison_AMPM}}
	\caption{Comparisons between different AM-AM and AM-PM models.}\label{fig:comparison}
\end{figure}

\subsection{Modified Rapp amplifier model}
\label{sec: Rapp}
The modified Rapp model mimics the performance of PAs that have a linear gain in the small signal region and a saturation for larger input values due to current and voltage clipping. Unlike the original Rapp model introduced in \cite{rapp1991effects}, the approach in \cite{IEEE80216m}, which is named the modified Rapp model, includes the effect of AM-PM distortion. The output power ($F_A$) and phase ($F_P$) are expressed as
%\begin{equation}
%    F_{AMAM} (x) = \frac {G_x}{(1 + | \frac {G_x}{V_{sat}}|^{2p})^{\frac{1}{2p}}},
%\end{equation}
%\begin{equation}
%    F_{AMPM} (x) = \frac {Ax^{q_1}} {1 + |\frac{x}{B}|^{q_2}},
%\end{equation}
\begin{align}
    F_{A} (x) = \frac {G x}{(1 + | \frac {G x}{V_{\mathrm{sat}}}|^{2p})^{\frac{1}{2p}}}, \; & \;
    F_{P} (x) = \frac {Ax^{q_1}} {1 + |\frac{x}{B}|^{q_2}},
    \label{eq:Rapp model}
\end{align}
where $G$ represents the small signal gain, $p$ denotes the smoothness factor, $V_{\mathrm{sat}}$ is the saturation voltage coefficient, and the coefficients $A$, $B$, $q_1$, $q_2$ are AM-PM distortion curve parameters. %Tab.~\ref{tab:rapp_coeff} reports the modelling parameters obtained by fitting our measurements at \textcolor{blue}{$f_c=280$~GHz}. 
\textcolor{black}{To derive the modified Rapp parameters, we adopt the same approach as that defined in Section \ref{chapter_Modelling}.B for the Ghorbani model.}
The Rapp AM-AM and AM-PM coefficients for $f_c$\,=\,$315$\,GHz are: $G$\,=\,13.07, $V_{\mathrm{sat}}$\,=\,0.0559\,V, $p$\,=\,0.878, $A=-1.7204 \times 10^{5}$, $B=8.5695\times 10^{-3}$, $q_1$\,=\,1.6949, $q_2$\,=\,1.7404.
The results are depicted in Figs.\,\ref{fig:All_AMAM} and \ref{fig:All_AMPM}. Reference \cite{Model_data} provides the coefficients of the Rapp PA nonlinearity model for different center frequencies $f_c$.

\subsection{Frequency dependence}
\begin{figure}[t]
    \centering
    \includegraphics[width=0.8\columnwidth]{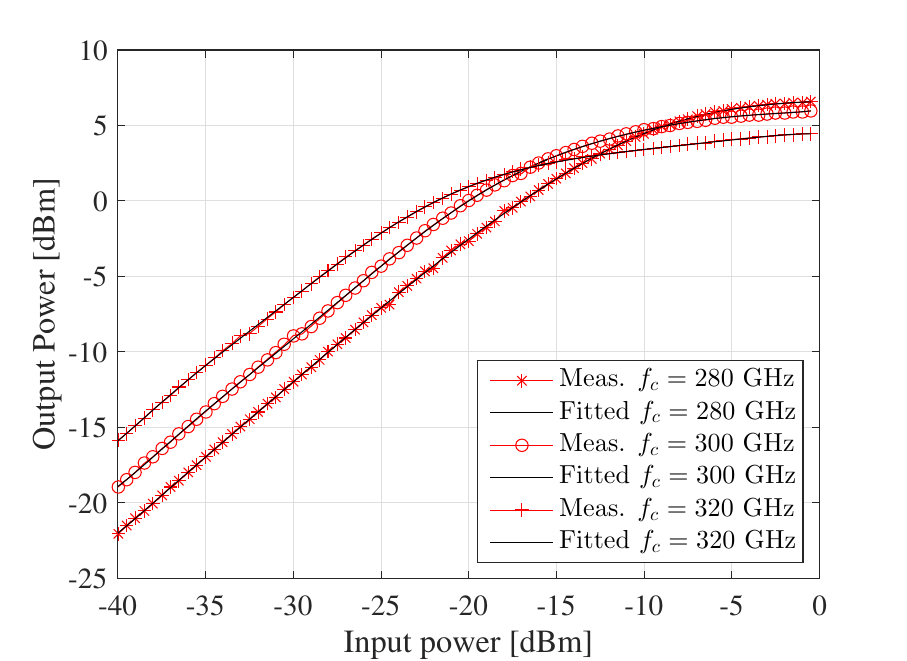}
    \caption{Different AM-AM measurement data and $9^{th}$ order polynomial behavioral models for different center frequencies.}
    \label{fig:diff_freqz_AMAM} % \vspace{-0.5cm}
\end{figure}

\begin{figure}[t]
    \centering
    \includegraphics[width=0.8\columnwidth]{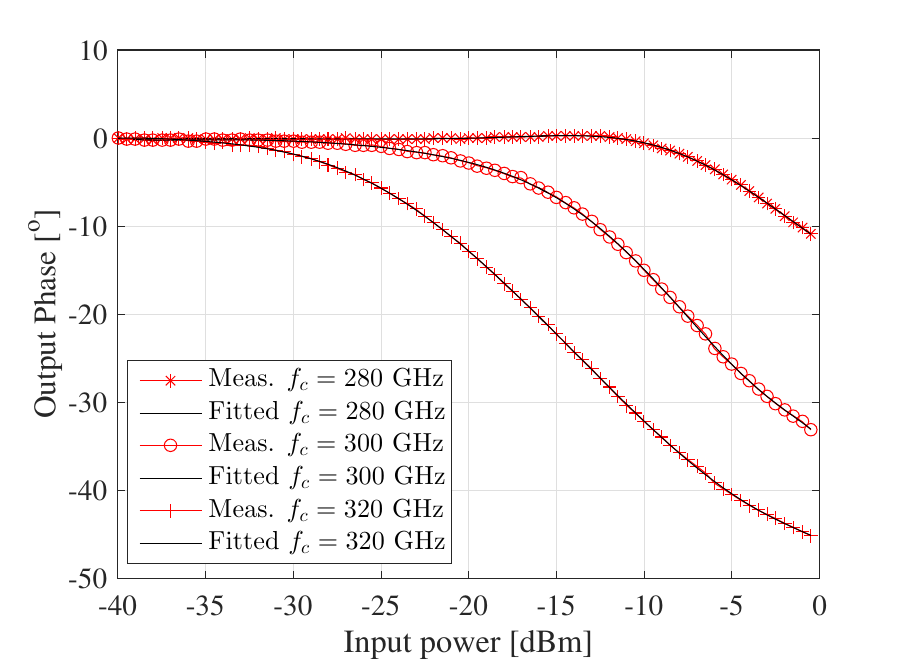}
    \caption{Different AM-PM measurement data and $9^{th}$ order polynomial behavioral models for different center frequencies.}
    \label{fig:diff_freqz_AMPM} % \vspace{-0.5cm}
\end{figure}

\begin{figure}[t]
    \centering
    \includegraphics[width=0.8\columnwidth]{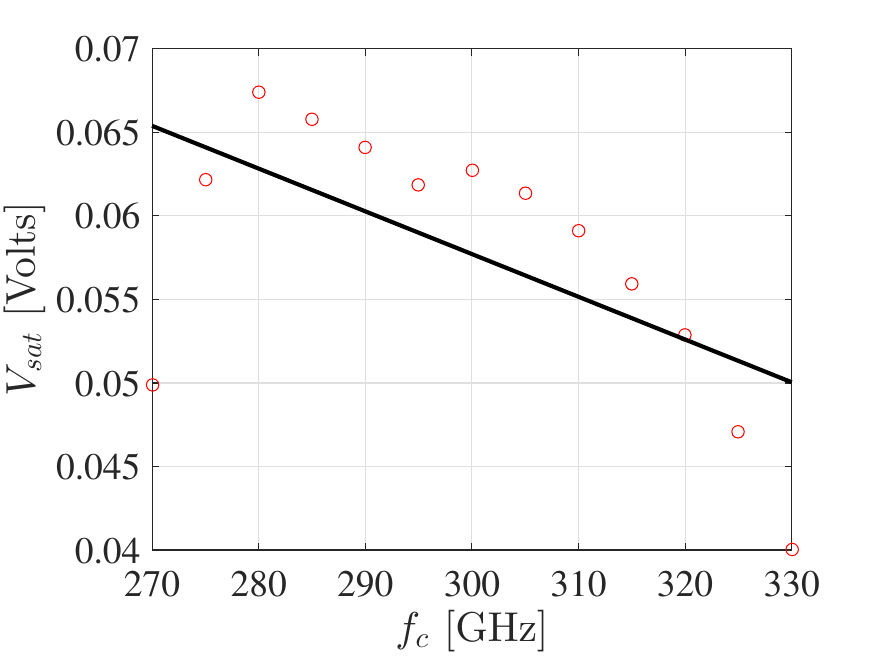}
    \caption{Rapp-based $V_{\mathrm{sat}}$ vs. operating frequency.}
    \label{fig:Vsat} % \vspace{-0.5cm}
\end{figure}

To capture the PA nonlinearity dependence on frequency, \textcolor{black}{we first depict in Figs. \ref{fig:diff_freqz_AMAM} and \ref{fig:diff_freqz_AMPM} the AM-AM and AM-PM measurements and polynomial behavioral models for various center frequencies, namely $f_c \in \{280,300,320\}$ GHz. As can be seen, there is a significant difference in the responses, emphasizing on the frequency dependence of the AM-AM and AM-PM behaviors for different center frequencies.} Moreover, Fig. \ref{fig:Vsat} depicts $V_{\mathrm{sat}}$ as a function of the operating frequency. The decreasing overall trend is captured in the figure, implying limitations on the transmission power as the frequency increases. This trend is also captured by the fitted first-order polynomial written as
\begin{align}
      V_{\mathrm{sat}}(f) = -2.5585\times 10^{-13} f + 0.1345,
\end{align}
where $f$ is the frequency in Hz. We note that the values of $V_{\mathrm{sat}}$ were extracted from the modified Rapp fitted parameters as a function of the operating frequency. The explainability of the AM-AM Rapp parameters is one of the main features that makes this model popular in standardization activities, while for the rest of the derived models, it is challenging to connect the parameters to the physical characteristics of the derived models, although some of them might yield a more accurate fit to the measurements.  

\subsection{Comparison with different models}
We now compare the derived model with models available in the literature as seen in Fig. \ref{fig:comparison}. The models that are selected to be compared with are:
\begin{itemize}
    \item 3GPP ($<6$ GHz): This is a polynomial-fitted model for both AM-AM and AM-PM, which was proposed in \cite{3gpp20205g2} for the purpose of evaluating waveforms operating below 6 GHz.
    \item 3GPP ($>6$ GHz): This model is a modified Rapp-fitted model, which was proposed in \cite{3gpp20205g2} to evaluate waveforms operating above 6 GHz, namely 30 and 70 GHz.
    \item SiGe-HEXA-X: SiGe-based modified Rapp model, which was proposed in \cite{singh2021290,HEXAXIID23} for an operating frequency of 290 GHz.
    \item ETSI ISG THz-based model, which is a modified Rapp model proposed in \cite{ISGTHzGR004} for operating frequencies in the 100-200 GHz band, extracted by averaging out the modified Rapp model parameters in the specified frequency range solely targeting AM-AM behavior. Parameters for AM-PM behavior were left unaddressed.
\end{itemize}
It is worth noting that the 3GPP ($<6$ GHz) model is strictly evaluated in the range of input powers $[-35,9]$ dBm, which is the range in which the polynomial was fitted, otherwise, the behavior of the model is unreliable. This hints to why the modified Rapp model is preferred over the fitted polynomial model, as the ranges of input power that are fed to the modified Rapp models exceed the range that it was fitted within while still producing an expectable behavioral performance. \textcolor{black}{One notable conclusion of this comparison is the over-estimation of the linear region of the PA models proposed by 3GPP (both higher and lower than 6 GHz), and ETSI ISG THz-based model (which omits the proposal of an AM-PM model) when compared to our derived model. For instance, at an input power value of $-10$ dBm, the 3GPP (both higher and lower than 6 GHz) and ETSI ISG THz-based models still operate in the linear region for the AM-AM case and introduce low phase shift values, while the {SiGe-based MMIC} and the modeled amplifier produce around $-30^{\circ}$ phase shift, and are clearly in the compression region, which is detrimental for even low-order QAM-based modulation.}

\subsection{Simulation and Comparison to Wideband Measurements}
\begin{figure}[]
	\centering
	\subfloat[]{\includegraphics[width=0.49\columnwidth]{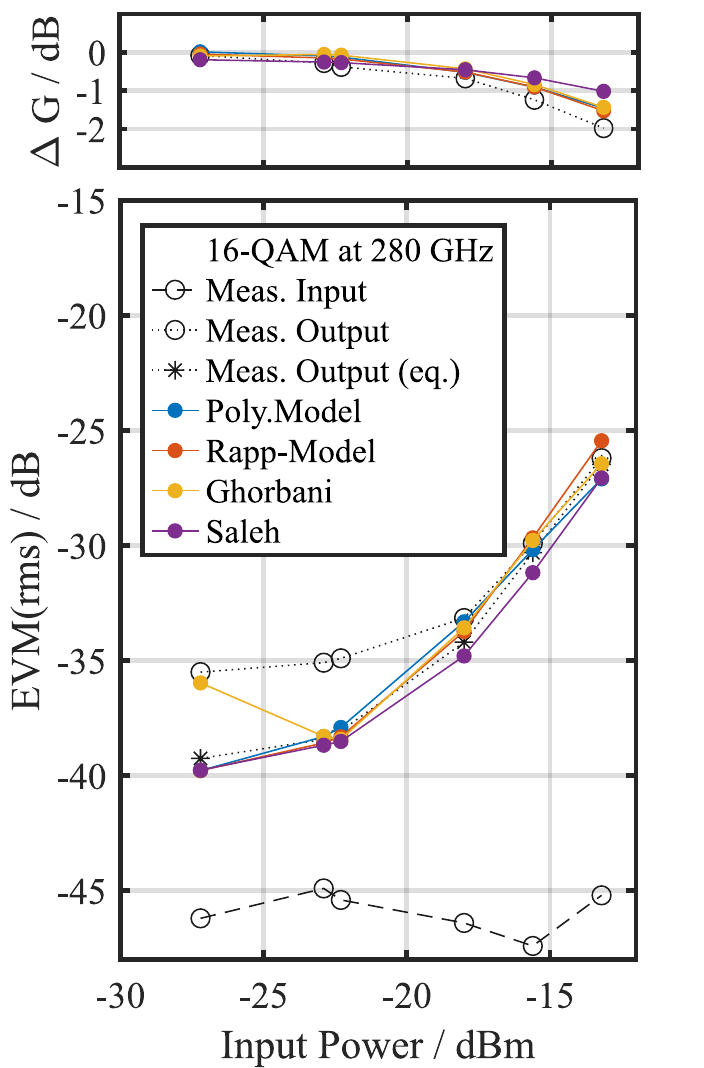}}
        \label{fig:Sim_EVM_16QAM}
	\subfloat[]
    {\includegraphics[width=0.49\columnwidth]{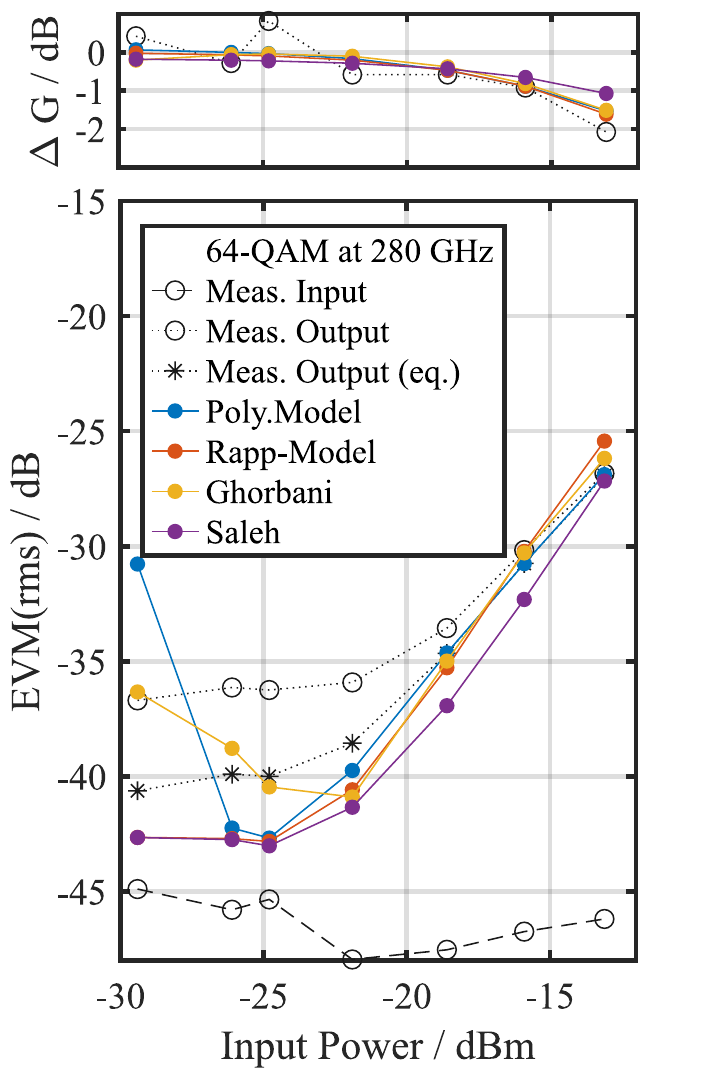}}
        \label{fig:Sim_EVM_64QAM}
    \\~\\
	\subfloat[]{\includegraphics[width=0.49\columnwidth]{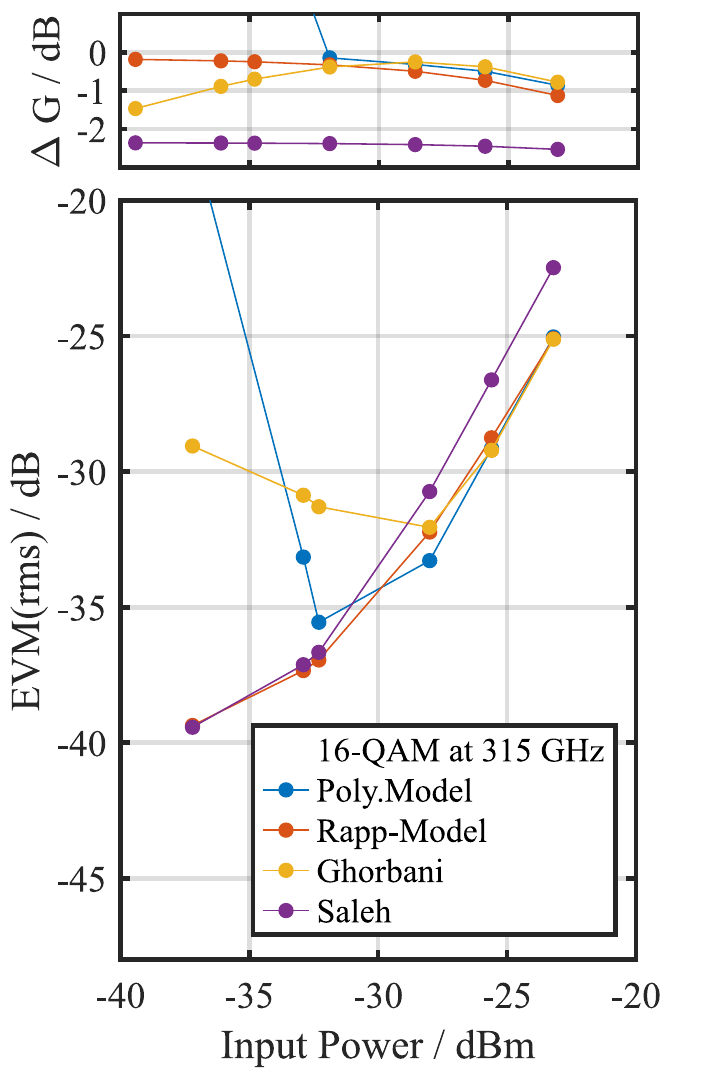}}
        \label{fig:Sim_EVM_16QAM_315GHz}
	\subfloat[]
    {\includegraphics[width=0.49\columnwidth]{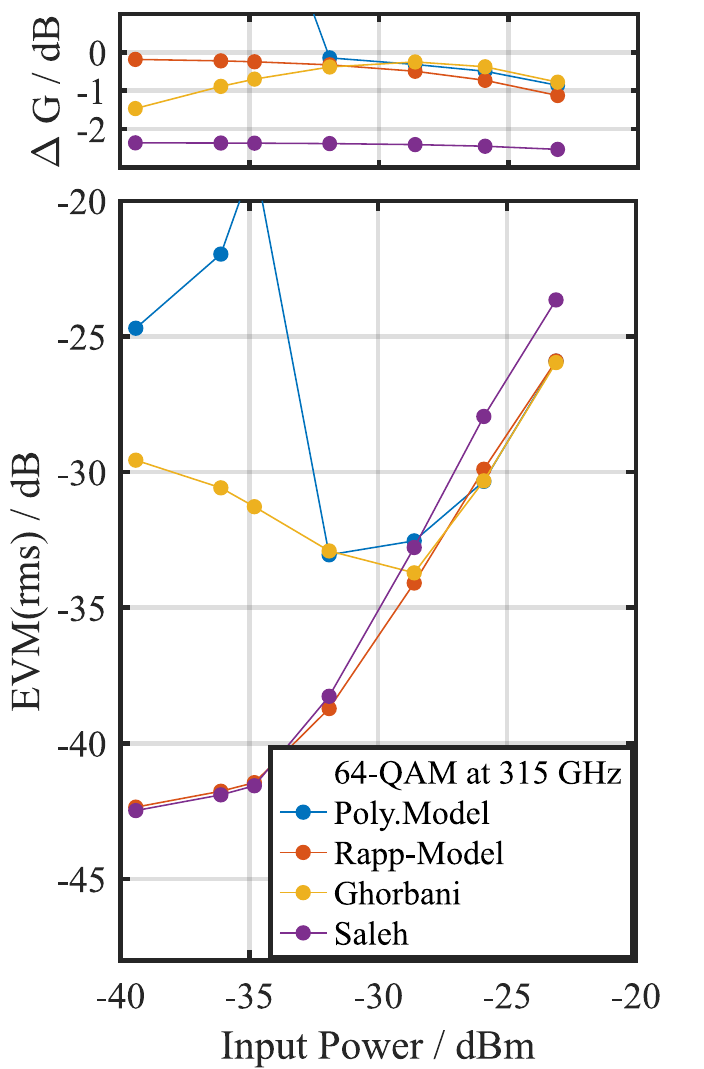}}
        \label{fig:Sim_EVM_64QAM_315GHz}
	%\hfil
	\caption{Comparison of the simulated \gls{evm} and gain-drop $\Delta G$ of the different non-linear models. (a) and (b) are showing 16-QAM and 64-QAM at 280\,GHz, supported by measured data. (c) and (d) are showing simulations for models extracted at 315\,GHz.}
	\label{fig:Sim_EVM}
\end{figure}

For the assessment of the model accuracy, the models are simulated for single-carrier QAM signals. For each model we consider two center frequencies, 280\,GHz and 315\,GHz. The results from the 280\,GHz simulation are further compared to the wideband measurements from Section \ref{ch:characterization}.
The simulated constellation diagrams at 280\,GHz, each for the lowest and highest input powers can be observed in Table~\ref{tab:SimulatedConstellations}. The Rapp model and the Saleh-model are showing similarly good results at all power levels. The polynomial model is showing inaccuracies for the lowest input powers. Although the \gls{rms} signal power lies within the definition range of the polynomial model, the instantaneous envelope power of the inner constellation points are exceeding the range at which the model has been extracted. Similarly, the Ghorbani model shows inaccurate modeling for lower input powers.
Fig.\,\ref{fig:Sim_EVM} shows a comparison of the \gls{evm} for all considered models and compares the simulated gain drop in the upper part of each plot. All models show a similar increase of \gls{evm} due to compression, which also aligns with the \gls{evm} measurement. However, in linear operation, polynomial and Ghorbani model are inaccurate. This issue is especially pronounced for higher order QAM-modulations like 64-QAM, because of the higher variance in amplitude states. For multi-carrier signals, this effect would be increased due to the even higher spread in amplitude states. For the 280\,GHz models, the gain decrease of the band power $\Delta G$ is captured equally good by all implementations, which is defined by applying
\begin{equation}
    \Delta G = P_{\mathrm{out}} - P_{\mathrm{in}} - S_{21}.
\end{equation}
With the difference of $P_{\mathrm{out}}$ and $P_{\mathrm{in}}$ being the simulated gain, normalized to the measured small-signal gain $S_{21}$. For the 315\,GHz models, only the Rapp model is able to reflect the gain compression correctly at all power levels.\\
We thus conclude that, out of the observed models, only the Rapp model should be considered for accurate system-level modeling. The modified Rapp model follows closely the measured \gls{evm} when linear equalization is applied, as shown in Fig.\,\ref{fig:Sim_EVM} (a) and \ref{fig:Sim_EVM} (b). This leads to the assumption, that PA distortion can be modeled accurately with memory-less PA-models in conjunction with linear memory-effects. 

% \textcolor{black}{Although the small signal region generally shows an underestimation of the EVM when it is computed using the developed models, a gradual agreement between the simulated and measured values is observed as the input power increases. Increasing the input power yields high output power values that are likely to be adopted \textit{to meet link budget demands}. Hence, we believe that the distortion produced by the derived quasi-memoryless models are of value to the communication community.}

Although the small signal region generally shows an underestimation of the EVM when it is computed using the developed models, a gradual agreement between the simulated and measured values is observed as the input power increases. Increasing the input power yields high output power values that are likely to be adopted \textit{to meet link budget requirements}. In other words, we expect the PA to operate mainly in proximity to the saturation region, where there is a good agreement between the measured distortions and those predicted by our quasi-memoryless models, which can then be used effectively for system performance analysis.

{
\begin{table}[t]
\caption{Comparison of the simulated ouput signals with the implemented PA-models at 280\,GHz.}
\begin{tabular}{|p{1.2cm}|p{2.9cm}|p{2.9cm}|}
\hline
     & \textbf{64-QAM simulated} & \textbf{64-QAM simulated}  \\
     & \textbf{with $P_{\mathrm{in}}$\,=\,-30\,dBm} & \textbf{with $P_{\mathrm{in}}$\,=\,-13.2\,dBm} \\ \hline
Polynomial model 
& \raisebox{-0.4\totalheight}{\includegraphics[width=2.6cm]{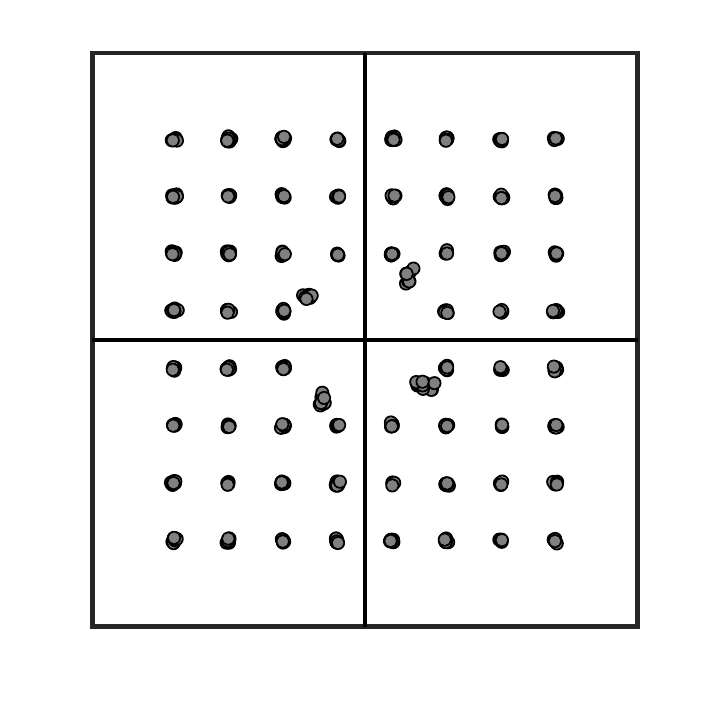}} \newline EVM\,=\,-30.8\,dB \newline $P_{\mathrm{out}}$\,=\,-11.4\,dBm
& \raisebox{-0.4\totalheight}{\includegraphics[width=2.6cm]{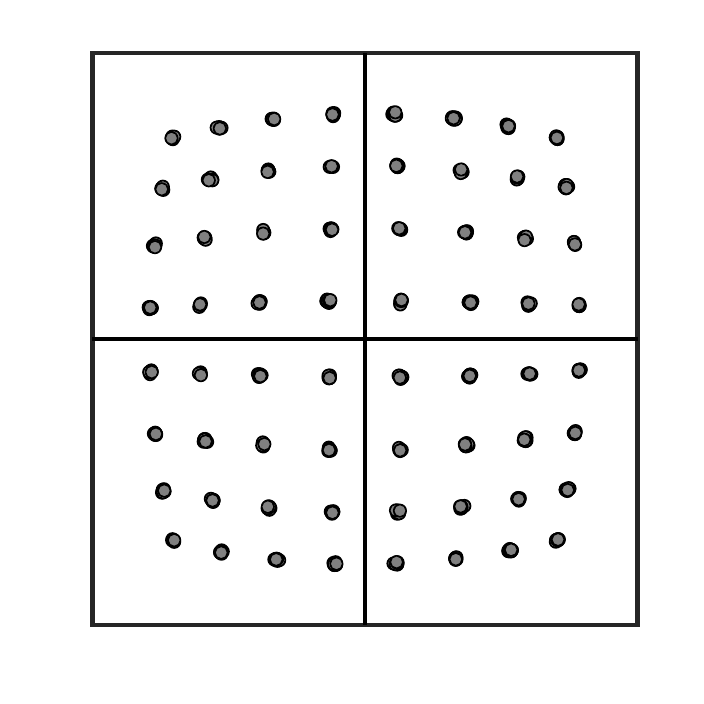}} \newline EVM\,=\,-26.9\,dB \newline $P_{\mathrm{out}}$\,=\,3.3\,dBm
\\ \hline
Rapp model 
& \raisebox{-0.4\totalheight}{\includegraphics[width=2.6cm]{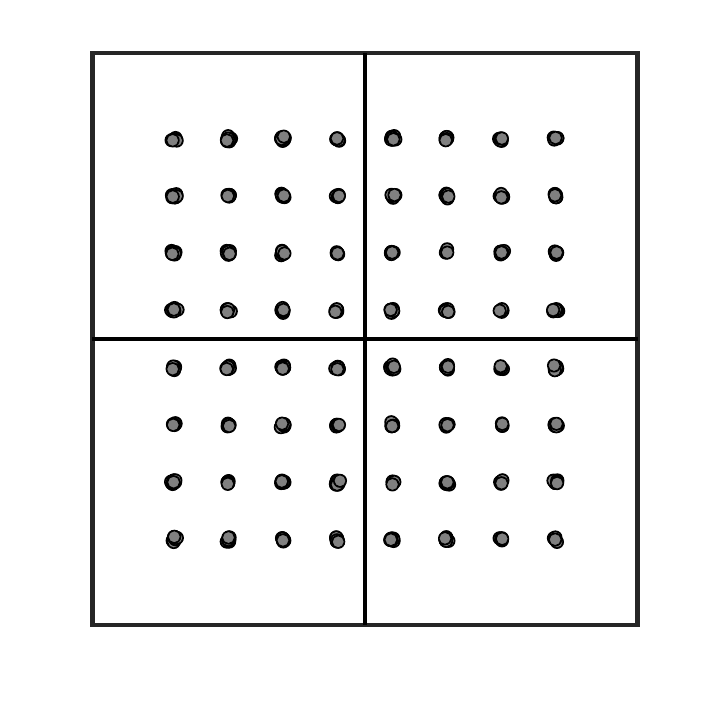}} \newline EVM\,=\,-42.7\,dB \newline $P_{\mathrm{out}}$\,=\,-10.9\,dBm
& \raisebox{-0.4\totalheight}{\includegraphics[width=2.6cm]{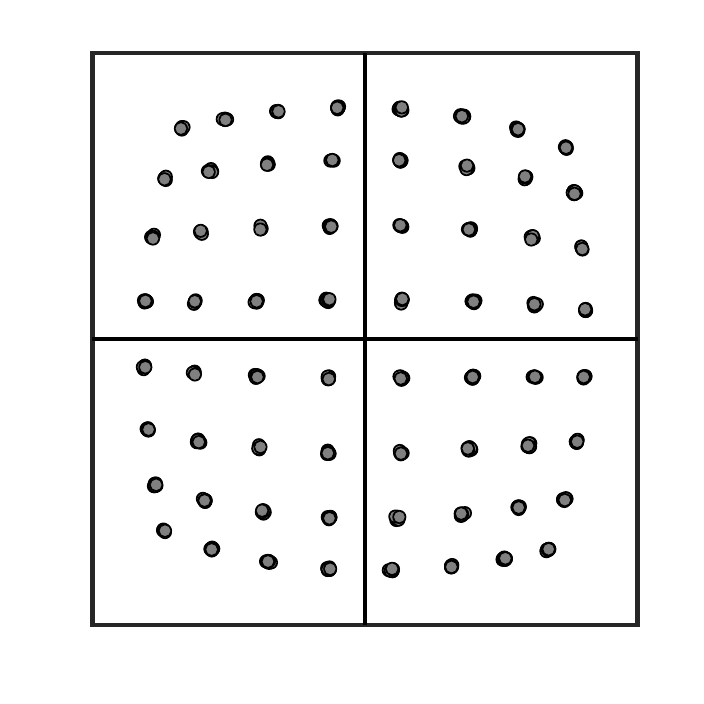}} \newline EVM\,=\,-25.4\,dB \newline $P_{\mathrm{out}}$\,=\,3.8\,dBm
\\ \hline
Ghorbani model 
& \raisebox{-0.4\totalheight}{\includegraphics[width=2.6cm]{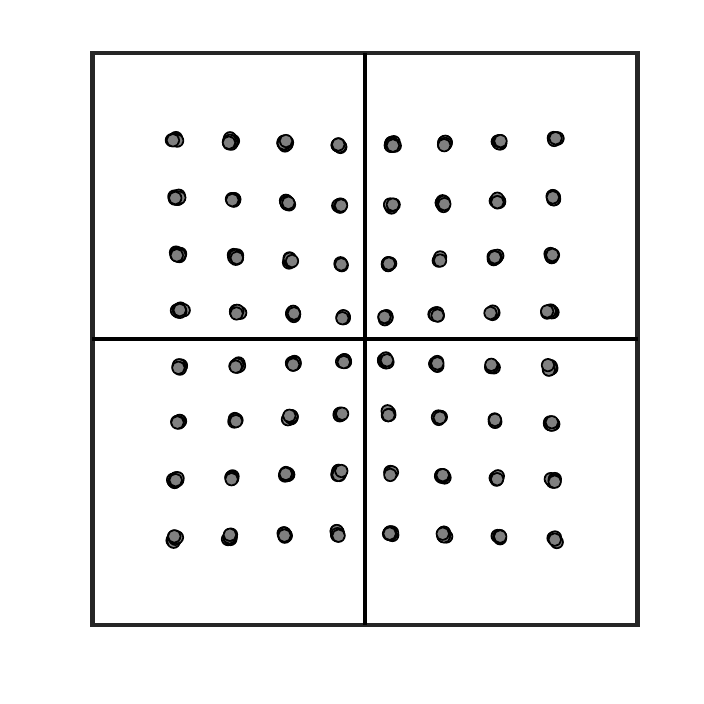}} \newline EVM\,=\,-34\,dB \newline $P_{\mathrm{out}}$\,=\,-11.4\,dBm
& \raisebox{-0.4\totalheight}{\includegraphics[width=2.6cm]{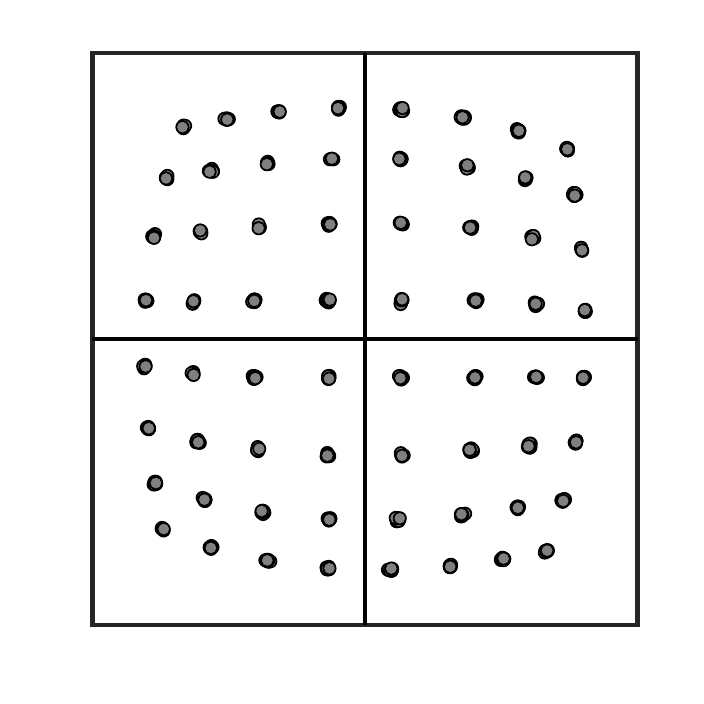}} \newline EVM\,=\,-25.0\,dB \newline $P_{\mathrm{out}}$\,=\,3.6\,dBm
\\ \hline
Saleh model 
& \raisebox{-0.4\totalheight}{\includegraphics[width=2.6cm]{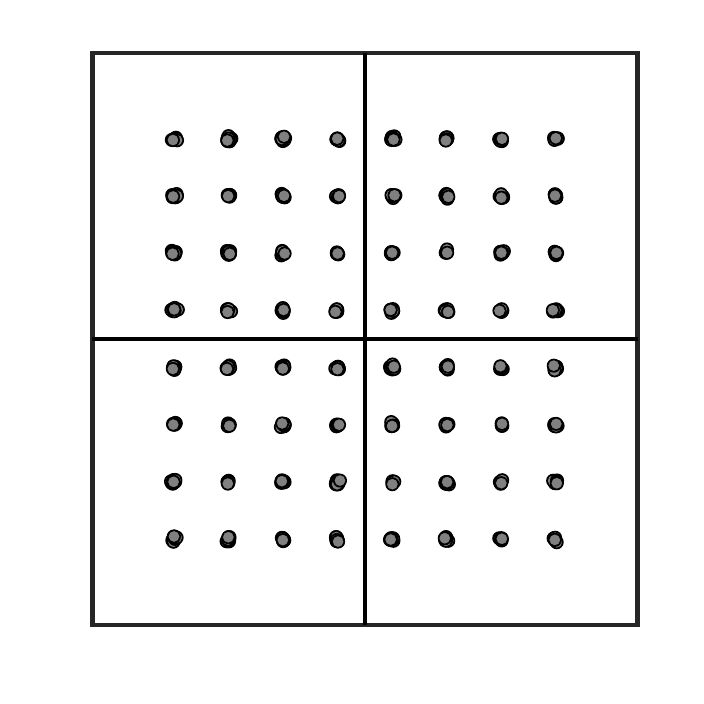}} \newline EVM\,=\,-42.7\,dB \newline $P_{\mathrm{out}}$\,=\,-11.6\,dBm
& \raisebox{-0.4\totalheight}{\includegraphics[width=2.6cm]{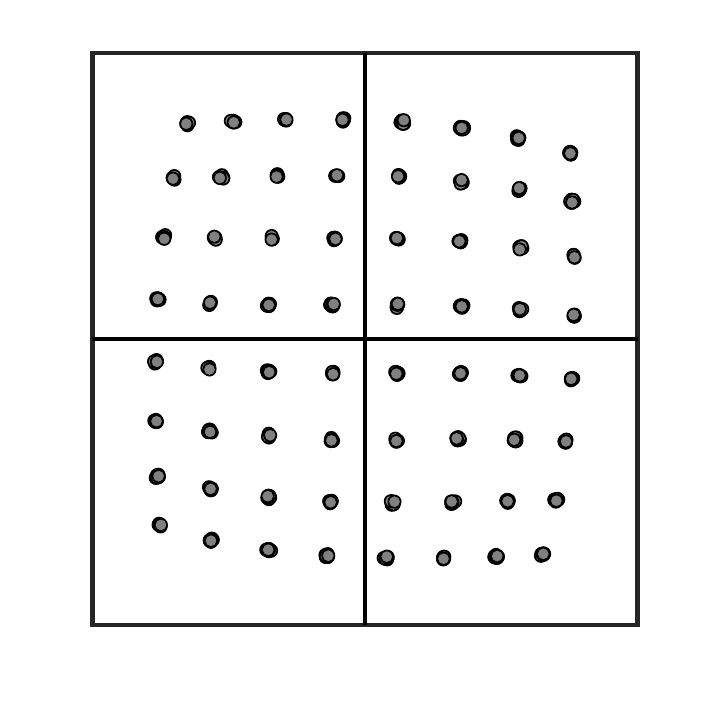}} \newline EVM\,=\,-27.2\,dB \newline $P_{\mathrm{out}}$\,=\,3.8\,dBm
\\ \hline
\end{tabular}
\label{tab:SimulatedConstellations}
%\vspace{-\baselineskip}% remove one line of space below this table
\end{table}
}

\section{Predistortion Algorithm}
\label{sec: PD}

In this section we describe and analyze the predistortion algorithm used to mitigate the non-linear effects of the PA. We start with the AM-AM and AM-PM characteristics provided by the modified Rapp model (as in \eqref{eq:Rapp model}) based on the discussions in Section~\ref{chapter_Modelling}, and derive the ideal predistortion function necessary to achieve perfect linearization. Specifically, we determine the inverse AM-AM characteristic and the required AM-PM compensation.  Practical implementation of the predistorter is then addressed, focusing on polynomial approximations of the ideal predistortion functions.

The signal $x(t)$ at the input of the predistorter in Fig.~\ref{fig: SystemModel} is a two-dimensional signal, $x(t) = \left(x_I(t),x_Q(t)\right)$ where $x_I(t)$ and $x_Q(t)$ are the In-phase and Quadrature components as usual. For mathematical convenience, $x(t)$ will be represented as a complex signal $\tilde{x}(t) = x_I(t) + {\rm j} x_Q(t)$. Similarly, $y(t) = \left(y_I(t),y_Q(t)\right)$ will be described through $\tilde{y}(t) = y_I(t) + {\rm j} y_Q(t)$. The signal at the output of the \gls{rf} converter can be written as
\begin{equation}
\label{eq: yRF}
\begin{split}
y_{\rm RF}(t) &= y_I(t) \cos(2 \pi f_c t) - y_Q(t) \sin(2 \pi f_c t), \\
&= \rho_y(t) \cos ( 2 \pi f_c t + \theta_y (t)),
\end{split}
\end{equation}
where 
\begin{equation}
\label{eq: roYtetaY}
\rho_y(t) = \sqrt{y^2_I(t)+ y^2_Q(t)} \quad \quad \theta_y (t) =\tan^{-1}\left( \dfrac{y_Q(t)}{y_I(t)}\right).
\end{equation}
Based on \eqref{eq:Rapp model}, the output of \gls{pa} is given by 
\begin{equation}
\label{eq: OutputHPA}
z(t) = F_{A} (\rho_y(t)) \cos ( 2 \pi f_c t + \theta_y (t) + \phi(t)),
\end{equation}
with $\phi(t) = F_{P} (\rho_y(t))$. From \eqref{eq: OutputHPA}, it is clear that the output amplitude and phase distortion, $F_{A} (\rho_y(t))$ and $F_{P} (\rho_y(t))$, respectively, depend only on $\rho_y(t) = |\tilde{y}(t)|$. This means that the predistortion algorithm can effectively operate on the baseband signals $\tilde{x}(t)$ and $\tilde{y}(t)$.

A key assumption made here is that \gls{pa} is a memoryless device, meaning that the output is only a function of the instantaneous input. Accordingly, for the sake of simplicity, we drop the dependance on time from all the signals. 

%\begin{figure}[ht]
%    \centering
%    \includegraphics[width=0.9\linewidth]{Figures/Predistortion_Algorithm/HPA-PD_block_scheme.jpg}
%    \caption{Predistortion-PA block scheme}
%    \label{fig:pa-pd-block-scheme}
%\end{figure}

\subsection{Ideal Predistortion Functions}

\begin{figure}[t!]
\begin{center}
\begin{tikzpicture}[scale=0.4, every node/.style={scale=0.4}]
[every text node part/.style={align=center}]

\draw[{}-{Stealth[length=2mm,width=2mm]}] (0,0) -- (1.5,0);
\node[above] at (0.75,0.2) {\LARGE${\tilde{x}(t)}$};

\draw[thick]  (1.5,-1) rectangle (4.5,1);
\node[align = center] at (3.0,0) {\LARGE$\textsf{Modulus}$};

\draw[{}-{Stealth[length=2mm,width=2mm]}] (4.5,0) -- (6,0);
\node[above] at (5.25,0.2) {\LARGE$\rho_x(t)$};

\draw (0.5,0) -- (0.5,-4);
\draw[{}-{Stealth[length=2mm,width=2mm]}] (0.5,-4) -- (1.5,-4);

\draw[thick]  (1.5,-5) rectangle (4.5,-3);
\node[align = center] at (3.0,-4) {\LARGE$\textsf{Arg}$};

\draw (4.5,-4) -- (15.0,-4);
\draw[{}-{Stealth[length=2mm,width=2mm]}] (15.0,-4) -- (15.0,-2.5);

\node[above] at (6,-3.8) {\LARGE$\theta_x(t)$};

\draw[thick]  (6,-1) rectangle (9,1);
\node[align = center] at (7.5,0) {\LARGE$A_{PD} (\cdot)$};

\draw[{}-{Stealth[length=2mm,width=2mm]}] (9,0) -- (19.5,0);
\node[above] at (10.5,0.2) {\LARGE$\rho_y(t)$};

\draw[thick] (20,0) circle (0.5);
\draw[thick][rotate around={45:(20,0)}](19.7,0) -- (20.3,0); 
\draw[thick][rotate around={-45:(20,0)}] (19.7,0) -- (20.3,0);

\draw[{}-{Stealth[length=2mm,width=2mm]}] (20.5,0) -- (22,0);
\node[above] at (21.25,0.2) {\LARGE$\tilde{y}(t)$};

\draw (9.5,0) -- (9.5,-2);
\draw[{}-{Stealth[length=2mm,width=2mm]}] (9.5,-2) -- (10.5,-2);

\draw[thick]  (10.5,-3) rectangle (13.5,-1);
\node[align = center] at (12.0,-2) {\LARGE$\Theta(\cdot)$};

\draw[{}-{Stealth[length=2mm,width=2mm]}] (13.5,-2) -- (14.5,-2);

\draw[thick] (15,-2) circle (0.5);
\draw[thick] (14.7,-2) -- (15.3,-2); 
\draw[thick][rotate around={90:(15,-2)}] (14.7,-2) -- (15.3,-2);

\draw[{}-{Stealth[length=2mm,width=2mm]}] (15.5,-2) -- (16.5,-2);

\draw[thick]  (16.5,-3) rectangle (19.5,-1);
\node[align = center] at (18.0,-2) {\LARGE${\rm exp}\{j(\cdot)\}$};

\draw (19.5,-2) -- (20,-2);
\draw[{}-{Stealth[length=2mm,width=2mm]}] (20,-2) -- (20,-0.5);

\end{tikzpicture}
\end{center}
\caption{Schematic diagram of the predistortion device}
\label{fig:Predistortion algorithm block scheme}
\end{figure}
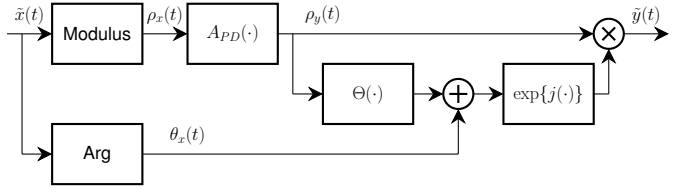

We start by representing $z$ with its complex envelope $\tilde z$, i.e.,
\begin{equation}
\label{eq: CEz}
\tilde z = \rho_z e^{{\rm j} \theta_z}
\end{equation}
where 
\begin{equation}
\label{eq: RoZTetaZ}
\rho_z = F_{A} (\rho_y) \quad \quad \theta_z = \theta_y + F_{P} (\rho_y).
\end{equation}
Ideally, we want the \gls{pa} output \(z\) to satisfy
\begin{equation}
    z = G x,
\end{equation}
or, equivalently,
\begin{equation}
    \begin{cases}
        \rho_z = G \rho_x \\
        \theta_z = \theta_x,
    \end{cases}
\label{eq: IdealPred}
\end{equation}
with $\rho_x = \sqrt{x^2_I + x^2_Q}$ and $\theta_x = \tan^{-1} (x_Q / x_I)$. Let \(A_{PD}(\rho)\) be the \textcolor{black}{amplitude function} of the predistorter, i.e., 
\begin{equation}
        \rho_y = A_{PD}(\rho_x),
\end{equation}
where $\rho_x$ and $\rho_y$ are the amplitudes at the input and output of \gls{pd}, respectively. To achieve perfect linearization of the \gls{pa}, the amplitude input-output response of the predistorter must be the inverse of the \gls{pa}'s response, i.e.,
\begin{equation}
\label{eq: IdealAmpPred}
    F_A[A_{PD}(\rho_x)] = G \rho_x.
\end{equation}
Denoting by \(F_A^{-1}\) the inverse of \(F_A(\cdot)\), we have
\begin{equation}
    A_{PD}(\rho_x) = F_A^{-1}[G \rho_x].
\end{equation}
The inverse of \(F_A(\cdot)\) can be computed as
\begin{equation}
    F_A^{-1}(G \rho_x) =
        \frac{\rho_x}{ \left[1 - \left(\frac{G \rho_x}{V_{\mathrm{sat}}}\right)^{2p} \right]^\frac{1}{2p} }
    \label{eq: PD amplitude function}.
\end{equation}
As for phase predistortion, based on \eqref{eq: RoZTetaZ} and \eqref{eq: IdealPred} we write
%\begin{equation}
%    \theta_x = \theta_y + F_{P} (\rho_y)
%\end{equation}
%or, equivalently,
\begin{equation}
    \theta_y = \theta_x - F_{P}(\rho_y).
\end{equation}
Since $\theta_x$ can easily be obtained by the input signal $x$, we just need to compensate for $F_{P}(\rho_y)$ for implementing an ideal phase predistortion. Accordingly, $\theta_y$ is generated as
\begin{equation}
\label{eq: IdealPhPred}
\theta_y =  \theta_x + \Theta(\rho_y),
\end{equation}
where  $\Theta(\rho) = - F_{P}(\rho)$ is the \textcolor{black}{ideal phase predistortion function}. A schematic diagram of \gls{pd} is reported in  Fig.~\ref{fig:Predistortion algorithm block scheme}. 
It is worth noting that, due to the discontinuity in the inverse function $F_A^{-1}(\cdot)$ at $\rho_x = V_{sat}/G$ (as can be seen from \eqref{eq: PD amplitude function}), the signal amplitude at the input of \gls{pd} must be limited through clipping. We denote by $\chi$ the clipping level at the input of \gls{pd} and by $\gamma$ the corresponding clipping value at the output, i.e., $\gamma =  A_{PD}(\chi)$. The impact of $\chi$ on system performance will be analyzed in detail in Section \ref{sec:Simulation Results}.

\subsection{Approximations of the ideal predistortion functions}
The ideal amplitude and phase functions of the predistorter, i.e., $A_{PD}(\rho_x)$ and $\Theta(\rho_y)$, can be approximated by polynomial models \cite{Dandrea1996} defined as
\begin{equation}
    A_{PD,{\rm poly}}(\rho) = \sum_{k=0}^{N_{A}} \eta_k \rho^k, \ \Theta_{{\rm poly}}(\rho) = \sum_{k=0}^{N_{\theta}} \nu_k \rho^k.
\end{equation}

%\begin{equation}
%    \Theta_{{\rm poly}}(\rho) = \sum_{k=0}^{N_{\theta}} \nu_k \rho^k 
%\end{equation}

The coefficients \(\eta_k\) and $\nu_k$ are obtained through a least-squares (LS) approach, by minimizing the following objective functions:
\begin{equation}
    \int\limits_{0}^{\chi} \left[ A_{PD}(\rho) - \sum_{k=0}^{N_{A}} \eta_k \rho^k \right] ^2 d\rho ,
\end{equation}

\begin{equation}
    \int\limits_{0}^{\gamma} \left[ \Theta(\rho) - \sum_{k=0}^{N_{\theta}} \nu_k \rho^k \right] ^2 d\rho .
\end{equation}

The impact of $N_{A}$ and $N_{\theta}$ on the system performance will be discussed in the next section.

%Once the polynomial coefficients are obtained, the predistortion algorithm can be implemented with the scheme represented in Fig.~\ref{fig:Predistortion algorithm block scheme}.

% \begin{figure}[ht]
%    \centering
%    \includegraphics[width=1\linewidth]{Figures/Predistortion_Algorithm/pd_block_scheme.png}
%    \caption{Predistortion algorithm block scheme}
%    \label{fig:Predistortion algorithm block scheme}
% \end{figure}

\section{Simulation Results}
\label{sec:Simulation Results}

Here we analyze the impact of the design parameters $\chi$, $N_{A} $ and $N_{\theta}$ on system performance, and discuss the possible trade offs between performance and predistortion algorithm complexity. Throughout this section, we will consider the communication system depicted in Fig.~\ref{fig: SystemModel}, and  we will assume that the input symbols belong to an $M$-QAM constellation, with $M=64$, the modulator uses a square-root raised-cosine filter \cite[Chap. 9]{Proakis2007} with a rolloff factor of $0.5$, and the carrier frequency is $f_c = 315$ GHz.

\begin{figure}[ht]
    \centering
    \includegraphics[width=0.7\linewidth]{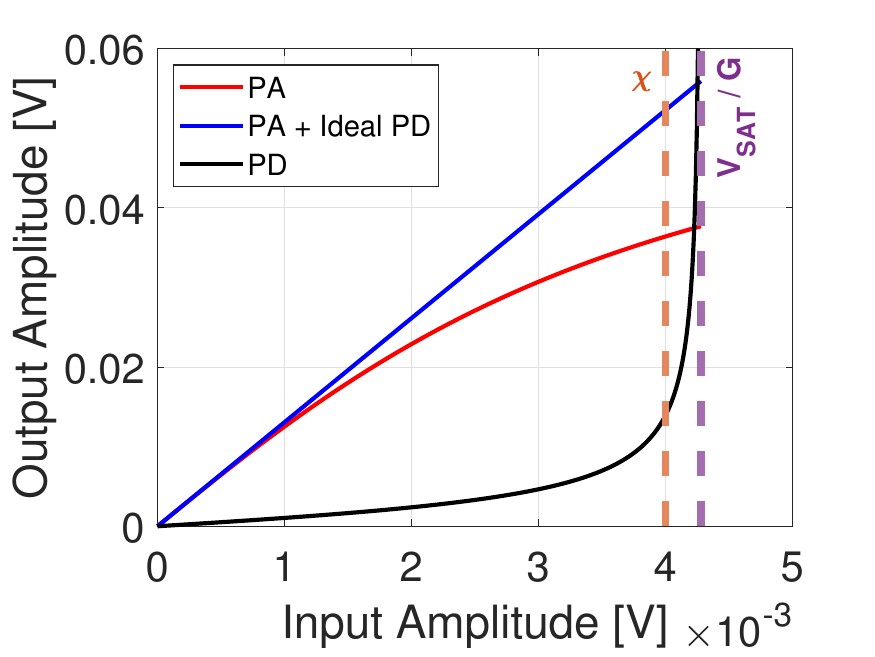}
    \caption{Output amplitude comparison between PA, PD, PA+PD}
    \label{fig:clipping_effects}
\end{figure}

\paragraph{Clipping}
Fig~\ref{fig:clipping_effects} shows the amplitude of the signals at the \gls{pa} output with no predistortion algorithm (red curve), at the output of the predistortion block (black curve), and at the output of \gls{pa} in the presence of an ideal predistorter (blue curve), as a function of the amplitude of the input signal $x(t)$. Looking at the black curve (\gls{pd} output), we can observe that, as the input amplitude approaches the $V_{sat}/G$ threshold (marked by the purple dashed line), the amplitude of $y(t)$ increases dramatically, as expected from \eqref{eq: PD amplitude function}. In order to keep $|y(t)|$ within reasonable values, the input amplitude $|x(t)|$ must be clipped at a level below the threshold $V_{sat}/G$, while still providing effective linearization of the \gls{pa}'s response. From Section~\ref{sec: Rapp}, we have $V_{sat}=0.0559$ V and $G=13.0732$, so that the threshold is about $4.28\times10^{-3}$~V.

%Allowing the input signal to reach this threshold would be problematic, as the extreme amplification at the predistorter output exceeds the PA output.
%To maintain it under control, the PD input amplitude should be clipped before reaching the exponential growth of the PD output amplitude. This ensures that the predistortion output amplitude remains within a manageable range, while still providing effective linearization of the power amplifier's response. 

Comparing the blue curve (\gls{pa} + Ideal \gls{pd}) to the original \gls{pa} response (red curve), we can see that the predistortion algorithm provides a nearly linear behavior up to an input amplitude of about $4\times10^{-3}$ V. The improvement is particularly evident in the mid-range region (between $1.5 \times 10^{-3}$ V and $3.5 \times 10^{-3}$ V) where the original \gls{pa} shows a significant compression. Based on the results in Fig~\ref{fig:clipping_effects}, we fix the clipping level at $\chi = 4\times10^{-3}$ V. 
%The linearized response shows a more consistent slope throughout most of the input range, maintaining nearly linear behavior up to about $4\cdot10^-3$ input amplitude.

%In this region, the predistortion successfully compensates for the PA's nonlinearity, resulting in a more linear overall response (blue curve maintaining a steadier slope).

\paragraph{Polynomial Order}
The impact of the polynomial approximation order ($N_A$,$N_\theta$) is analyzed to evaluate performance and complexity trade-off. In doing so, we assume that the modulated waveform $x(t)$ corresponds to an \gls{ofdm} signal with $N_{sc} =256$ subcarriers. Our analysis is carried out for different values of the Input Back-Off (IBO), which is the difference, measured in decibels (dB), between the average power of $x(t)$ and the 1-dB compression point of the \gls{pa}.

% The 1 dB compression point is defined as the input power where the \gls{pa}'s gain drops by 1 dB from its linear region. Operating at a lower input power (higher IBO) reduces the probability of non-linear distortions caused by the \gls{pa}, which can significantly impact signal integrity.,

% The input symbols belong to an $M$-QAM constellation, with $M=64$, and the modulator uses a square-root raised-cosine filter \cite[Chap. 9]{Proakis2007} with a rolloff factor of $0.5$. Finally, the carrier frequency is $f_c = 315$ GHz.

\begin{figure}[h]
	\centering
	\subfloat[]{\includegraphics[width=0.5\columnwidth]{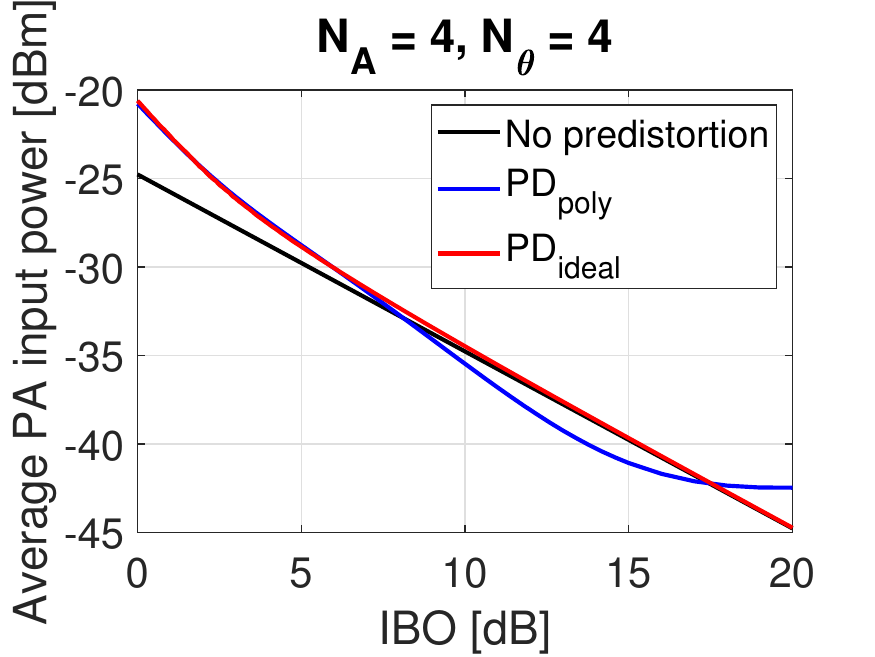}\label{fig:avg_pa_pin_N=4}}
	%\hfil
	\subfloat[]{\includegraphics[width=0.5\columnwidth]{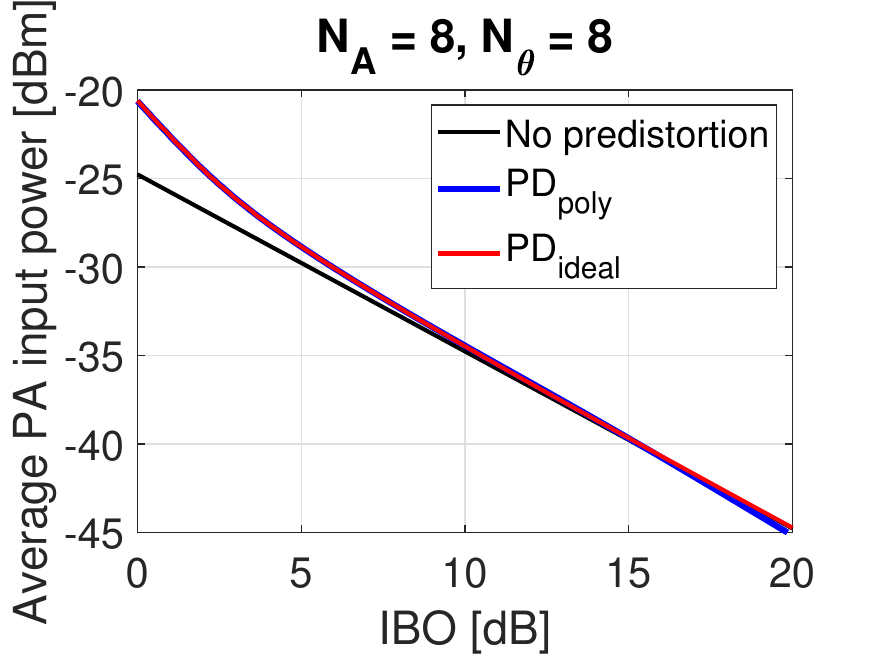}\label{fig:avg_pa_pin_N=8}}
	\caption{Average PA input power vs. IBO for different PD polynomial approximation orders.}\label{fig:avg_pa_pin}
\end{figure}

% Fig. \ref{fig:avg_pa_pin} shows the average power at the PA input as a function of IBO (\textcolor{black}{TO BE DEFINED}) for multiple system configurations:
% \begin{enumerate}
% \item Polynomial orders $N_A$ = $N_\theta$ = [4,8]
% \item Both predistorted and non-predistorted scenarios
% \item Number of subcarriers $N_{sc}$ = [256]
% \item Clipping thresholds $\chi$ = [0.002,0.003,0.004]
% \end{enumerate}
Fig. \ref{fig:avg_pa_pin_N=4} shows the average power at the \gls{pa} input as a function of IBO,
with a fourth-order polynomial approximation, i.e., $N_A = N_\theta = 4$.  For comparison, in the same figure we report the results without predistortion and for an ideal predistorter with the same clipping-level $\chi = 0.004$ V. As can be seen, at low IBO values the predistorted signal exhibits a higher average power compared to the non-predistorted case, meaning that the linearization of the \gls{pa} requires additional power when we are close to the compression point, as expected. In particular, the curve of the polynomial approximation is close to the curve of the ideal predistorter. As IBO increases, the curve of the polynomial predistorter shows considerable oscillations around the optimal value, meaning that, at high IBO values, a {fourth-order} polynomial approximation is not a good choice.

Similar behaviors can be observed with an eighth-order polynomial approximation, as shown in Fig.~\ref{fig:avg_pa_pin_N=8}. However, compared to the fourth-order case, we observe that the polynomial approximation and the ideal predistorter are very close, also at high IBO values. Accordingly, we expect to obtain almost ideal performance when using $N_A = N_\theta = 8$, at the price of increased complexity.

Although the results in Fig.~\ref{fig:avg_pa_pin} have been obtained with a multi-carrier signal characterized by $N_{sc} = 256$, the same conclusions can be drawn for different values of $N_{sc}$ and for single-carrier waveforms.

%The PA output power, Fig. \ref{fig:avg_pa_pout} exhibits similar characteristics to the input behavior shown in Fig. \ref{fig:avg_pa_pin}. At low IBOs, the power increase is more compressed due to operation near the PA's saturation region (1dB compression point). For $\chi$ = 0.004 with $N_A$ = $N_\theta$ = 4, the oscillatory behavior persists through the PA. These variations will significantly impact both EVM and BER performance, as examined in subsequent analyses.
%
%
%
%\begin{figure}[]
%	\centering
%	\subfloat[]{\includegraphics[width=0.5\columnwidth]{Figures/Simulation_Results/average_output_power_N=4.eps}\label{fig:avg_pa_pout_N=4}}
%	%\hfil
%	\subfloat[]{\includegraphics[width=0.5\columnwidth]{Figures/Simulation_Results/average_output_power_N=8.eps}\label{fig:avg_pa_pout_N=8}}
%	\caption{Average signal output power after PA}\label{fig:avg_pa_pout}
%\end{figure}

\paragraph{Error Vector Magnitude Analysis}

In this section, system performance in terms of \gls{evm} is evaluated for different configurations to assess the impact of the polynomial order. In particular, our analysis focuses on the behavior of \gls{evm} as a function of the number of subcarriers, providing important insights into system performance scalability. \gls{evm} has been evaluated by simulation, considering both single-carrier ($N_{sc} =1$) and multi-carrier signals, with $N_{sc} = \{16,32,64,128,256\}$. In all cases, the bandwidth is $B = 1\mathrm{\space GHz}$. For the power amplifier, two different values of IBO have been used, namely, $0$ and $10$ dB.

% EVM has been evaluated by simulation at the output of the receiver in Fig.~\ref{fig: SystemModel}, assuming an ideal propagation channel and neglecting the noise at the receiver. We consider both single-carrier ($N_{sc} =1$) and multi-carrier signals, with $N_{sc} = (16,32,64,128,256)$. In all cases, the bandwidth is $B = 1$ GHz. For the power amplifier, two different values of IBO have been used, namely, $0$ and $10$ dB.

% The main system parameters are listed in Table~\ref{tab:evm_system_params}.

% \begin{table}[h]
%     \centering
%     \begin{tabular}{c|c}
%         \hline
%         \textbf{Parameter} & \textbf{Value} \\ 
%         \hline
%         Number of subcarriers  & $N_{sc}$ = (1,16,32,...,256) \\ 
%         \hline
%         Modulation order  & $M = 64$\\ 
%         \hline
%         Oversampling rate & 10 \\ 
%         \hline
%         Rolloff factor  & $\beta_{rolloff}$ = 0.5 \\ 
%         \hline
%         Bandwidth  & $B$ = 1 GHz \\ 
%         \hline
%         Center frequency  & $f_c$ = 315 GHz \\ 
%         \hline
%         Input back-off & (0,10) dB \\ 
%         \hline
%     \end{tabular}
%     \caption{System Parameters}
%     \label{tab:evm_system_params}
% \end{table}

\begin{figure}[]
	\centering
	\subfloat[]{\includegraphics[width=0.5\columnwidth]{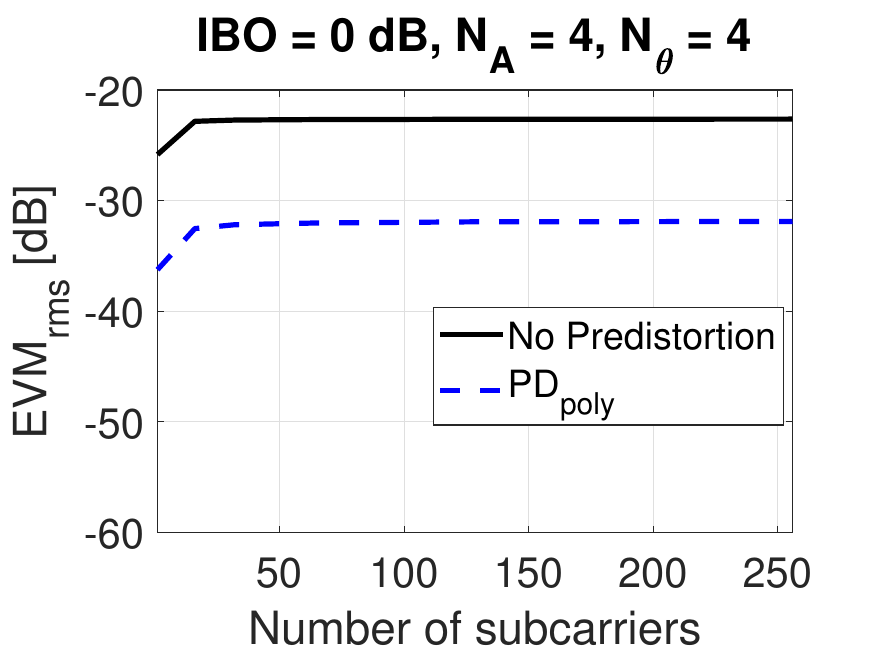}\label{fig:evm_vs_nsc_ibo0_N4}}
	%\hfil
	\subfloat[]{\includegraphics[width=0.5\columnwidth]{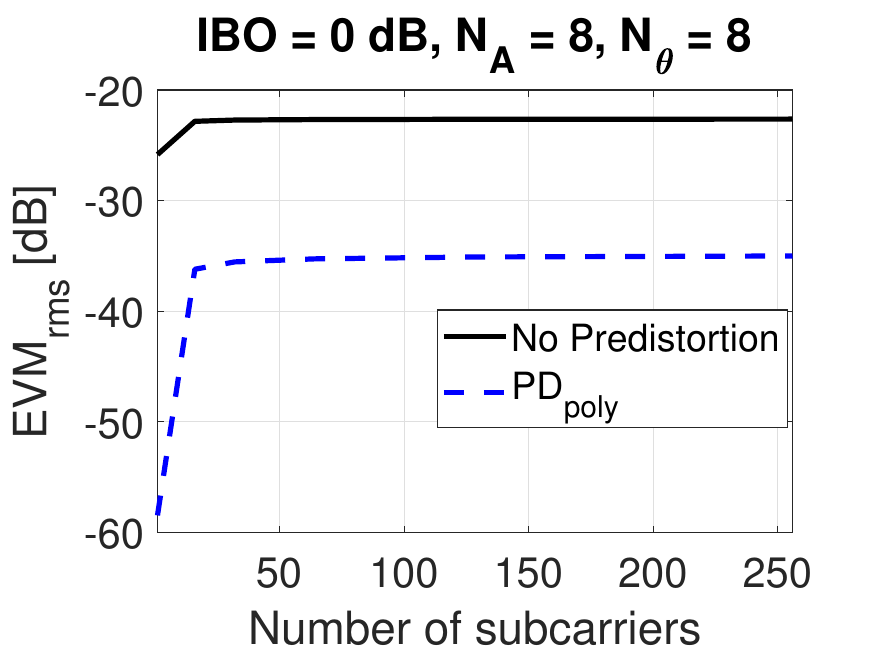}\label{fig:evm_vs_nsc_ibo0_N8}}
	\caption{EVM vs $N_{sc}$, IBO = 0 dB}\label{fig:evm_vs_nsc_ibo0}
\end{figure}

Fig.~\ref{fig:evm_vs_nsc_ibo0_N4} shows \gls{evm} as a function of the numbers of subcarriers, for IBO $= 0$ dB and $N_A = N_\theta = 4$. As can be seen, the predistortion algorithm improves the \gls{evm} values by approximately 10\,dB, compared to the case with no predistorter. This gain is the same for single-carrier and multi-carrier signals. It should be noted that the \gls{evm} values are almost insensitive to the number of subcarriers, except when we pass from $N_{sc} = 1$ to $N_{sc} = 16$. Similar conclusions can be drawn from the results in Fig.~\ref{fig:evm_vs_nsc_ibo0_N8} obtained IBO $= 0$ dB and $N_A = N_\theta = 8$. As expected, increasing the polynomial order provides better performance, with a gain of about 15\,dB in the multi-carrier case and 20\,dB with a single-carrier signal.

% In Fig.~\ref{fig:evm_vs_nsc_ibo0_N8}, we present the EVM performance for IBO = 0 dB and $N_A = N_\theta = 8$. A comparative analysis with the $N_A = N_\theta = 4$ configuration reveals several significant observations:
% \begin{itemize}
%     \item For $\chi = 0.002$ and $\chi = 0.003$, increasing the predistortion model complexity yields negligible performance improvements, with EVM values remaining practically identical to those observed in the lower-complexity configuration.
%     \item For $\chi = 0.004$, however, we observe substantial performance enhancements across both single-carrier and multi-carrier scenarios. In the single-carrier configuration, the EVM is approximately 20 dB lower than with $N_A = N_\theta = 4$, establishing this clipping threshold as the optimal choice even for single-carrier transmission—a significant departure from the results observed with lower model complexity.
%     While the multi-carrier scenario also demonstrates improvements but the magnitude of enhancement is less pronounced than in the single-carrier case.
% \end{itemize}

\begin{figure}[]
	\centering
	\subfloat[]{\includegraphics[width=0.5\columnwidth,height=3.3cm]{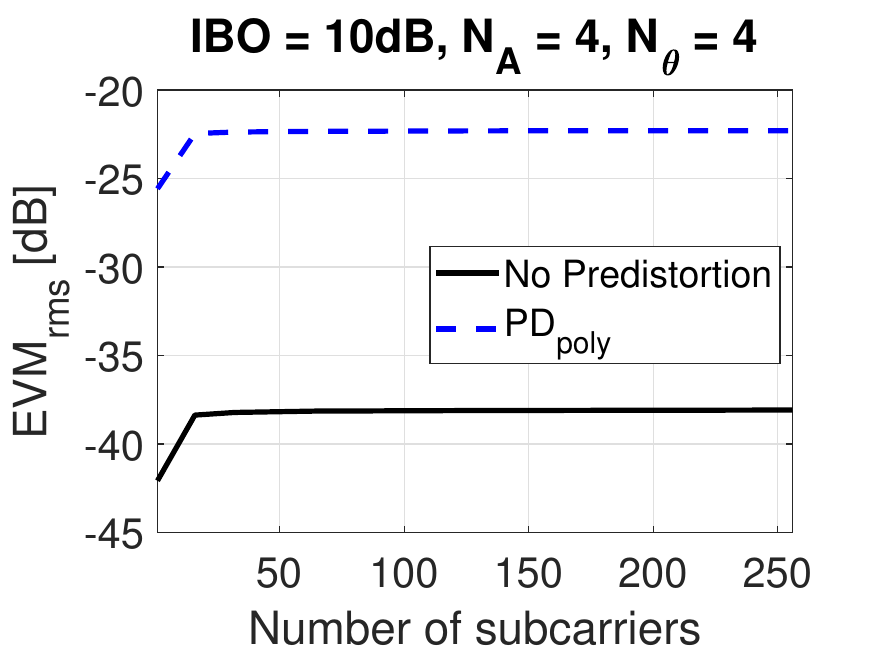}\label{fig:evm_vs_nsc_ibo10_N4}}
	%\hfil
	\subfloat[]{\includegraphics[width=0.5\columnwidth,height=3.3cm]{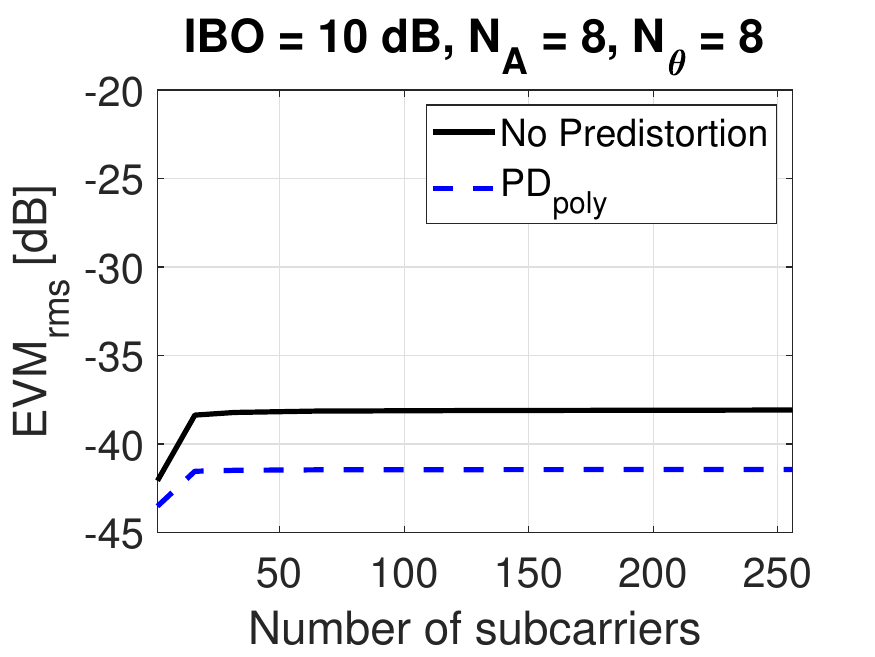}\label{fig:evm_vs_nsc_ibo10_N8}}
	\caption{EVM vs $N_{sc}$, IBO = 10 dB}\label{fig:evm_vs_nsc_ibo10}
\end{figure}

Fig.~\ref{fig:evm_vs_nsc_ibo10_N4} illustrates \gls{evm} as a function of $N_{sc}$, for IBO $= 10$ dB and $N_A = N_\theta = 4$. In this case, we observe a significant performance degradation compared to the scenario with no predistortion. This behavior is consistent with the results in Fig.~\ref{fig:avg_pa_pin_N=4} which show that, at high IBO values, the performance of the approximate predistorter is far from that of the ideal one. Passing from  $N_A = N_\theta = 4$ to $N_A = N_\theta = 8$ provides a substantial improvement, with \gls{evm} values that are now lower than the case of no predistortion, as can be seen from Fig.~\ref{fig:evm_vs_nsc_ibo10_N8}.

% In Fig.~\ref{fig:evm_vs_nsc_ibo10_N8}, we present the EVM performance for IBO = 10 dB and $N_A = N_\theta = 8$. When compared to the $N_A = N_\theta = 4$ configuration, several noteworthy trends emerge:

% \begin{itemize} 
% \item At $\chi = 0.004$, we observe a substantial performance improvement, with EVM values now lower than the non-predistorted case, though still not matching the performance achieved with the other clipping thresholds.
% \end{itemize}

% This comprehensive analysis underscores the critical importance of balancing predistortion parameters. Optimal performance at lower back-off values necessitates the use of higher clipping thresholds, albeit at the expense of increased computational complexity and degraded performance at higher back-offs. Conversely, superior performance at higher back-offs can be achieved with lower clipping thresholds, requiring reduced complexity but exhibiting diminished reliability near the power amplifier's saturation region.
% These EVM findings complement our power analysis results and establish a foundation for the subsequent BER performance evaluation, highlighting the intricate relationship between clipping threshold, polynomial order, and system performance across various operating conditions.
% As an ancillary observation unrelated to predistortion performance, we note that across all configurations, EVM degradation becomes negligible beyond $N_{sc} \geq 16$ subcarriers, suggesting that the primary performance impact manifests during the transition from single-carrier to multi-carrier transmission.

\paragraph{Bit Error Rate analysis}

The \gls{ber} of the communication system in Fig.~\ref{fig: SystemModel} has been analyzed considering a point-to-point link in which the transmit and receive antennas are at a distance $d$. The channel is AWGN and the received power $P_{r}$ is computed as
\begin{equation}
    P_{r} = P_{t} \dfrac{G_t G_r}{(4 \pi d/\lambda)^2 \cdot L(f_c,d)},
\end{equation}
where $P_{t}$ is the power of the transmitted signal, $G_t$ and $G_r$ are the transmit and receive antennas gains, respectively, $\lambda = c/f_c$ is the wavelength, $c$ being the speed of light in vacuum, and $L(f_c,d)$ is the atmospheric gaseous attenuation evaluated under standard conditions \textcolor{black}{({Atmospheric pressure}: 101.300\,kPa, {Temperature}: 290\,K, {Water vapor density}: 7.5\,g/{m$^3$})} as in \cite{ITU-R-P676-12}.
% \begin{itemize}
%     \item Atmospheric pressure: 101.300 kPa, 
%     \item Temperature: 290 K,
%     \item Water vapor density: 7.5 $\frac{g}{m^3}$,
% \end{itemize}

The noise power at the receiver has been computed by using the formula $\sigma^2 = k_{\rm B} T_n B$, where $k_{\rm B}$ is the Boltzmann constant and $T_n$ the equivalent noise temperature. Hence, the signal-to-noise ratio is given by SNR = $P_r/\sigma^2$.
The main system parameters are listed in Table~\ref{tab:ber_system_params}.

% \begin{table}[h]
%     \centering
%     \begin{tabular}{c|c}
%         \hline
%         \textbf{Parameter} & \textbf{Value} \\ 
%         \hline
%         Number of subcarriers  & $N_{sc}$ = \{1,256\} \\ 
%         \hline
%         Modulation order  & $M$ 64-QAM \\ 
%         \hline
%         Oversampling rate & 10 \\ 
%         \hline
%         Rolloff factor  & 0.5 \\ 
%         \hline
%         Transmit antenna gain  & $G_t$ = 45 dBi \\ 
%         \hline
%         Receive antenna gain  & $G_r$ = 14 dBi \\ 
%         \hline
%         Distance  & $d$ = 35 m \\ 
%         \hline
%         Bandwidth  & $B$ = 1 GHz \\ 
%         \hline
%         Noise temperature & T = 290 K \\
%         \hline
%         Center frequency  & $f_c$ = 315 GHz \\ 
%         \hline
%         Input back-off & \{0,1,..., 20\} dB \\ 
%         \hline
%     \end{tabular}
%     \caption{System Parameters}
%     \label{tab:ber_system_params}
% \end{table}

\begin{table}[h]
    \centering
    \begin{tabular}{c|c}
        \hline
        \textbf{Parameter} & \textbf{Value} \\ 
        \hline
        Number of subcarriers  & $N_{sc}$ = \{1,256\} \\ 
        \hline
        Modulation order  & $M = 16,\,64$ \\   
        \hline
        Transmit antenna gain  & $G_t$ = 45 dBi \\ 
        \hline
        Receive antenna gain  & $G_r$ = 14 dBi \\
        \hline
        Bandwidth  & $B$ = 1 GHz \\ 
        \hline
        Noise temperature & $T_n = 290$ K \\
        \hline
        Center frequency  & $f_c$ = 315 GHz \\ 
        \hline
    \end{tabular}
    \caption{System Parameters}
    \label{tab:ber_system_params}
\end{table}

\begin{figure}[]
	\centering
	\subfloat[]{\includegraphics[width=0.5\columnwidth, height=3.2cm]{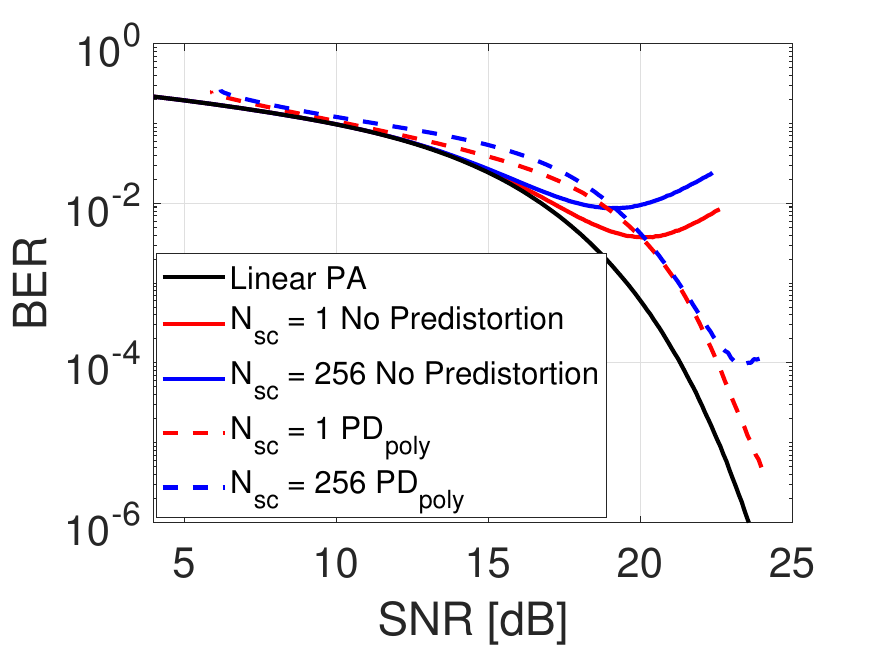}\label{fig:ber_vs_SNR_N=4}}
	%\hfil
	\subfloat[]{\includegraphics[width=0.5\columnwidth, height=3.2cm]{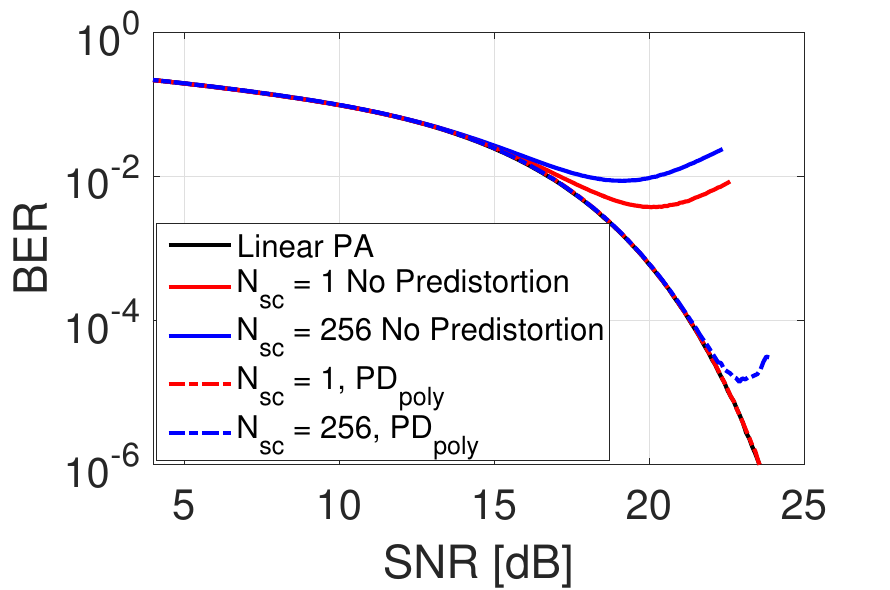}\label{fig:ber_vs_SNR_N=8}}
	\caption{BER vs SNR performance, 64 QAM. (a) $N_A=4,N_\theta=4$, (b) $N_A=8,N_\theta=8$}\label{fig:BERvsSNR}
\end{figure}

Fig. \ref{fig:ber_vs_SNR_N=4} shows BER as a function of SNR, with $N_{A}= N_{\theta}= 4$. The modulation order is $M=64$, and the distance between the transmitter and the receiver is $d=35$ m. The red curves pertain to the single-carrier case ($N_{sc}=1$) while the blue curves have been obtained with $N_{sc}=256$.   
% In Fig. \ref{fig:BERvsSNR} the \gls{ber} curves are plotted for $N_{A}= N_{\theta}= [4,8]$ The red curves represent the single carrier case and the blue curves represent the 256 subcarrier \gls{ofdm} case. 
The ideal BER, with a linear PA, and the BER without predistortion have been reported for comparison. It can be seen that, at low SNRs, corresponding to low transmit powers and, hence, to a \gls{pa} operating in the linear region, the performance without predistortion is virtually identical to that of the ideal system, and is better than that of the system with predistortion. This fact can easily be explained with the results in Fig.~\ref{fig:avg_pa_pin_N=4} which show that, at high IBO values, i.e., in the linear region of the \gls{pa}, the performance of the polynomial predistorter with $N_{A}= N_{\theta}= 4$ does not coincide with that of the ideal one. As the SNR increases (and IBO decreases) above $16$ dB, the BER of the system without predistortion initially decreases (but deviates from the ideal BER) until a minimum is reached, at $\rm{SNR} \approx 20$\,dB in the single-carrier case and $\rm{SNR} \approx 18$ dB for $N_{sc} = 256$. A further increase in the SNR results in a higher BER due to the uncompensated distortions produced by the \gls{pa}, which now operates in the non-linear region. Better results can be obtained with the predistortion algorithm. In particular, with $N_{sc} =1$, the predistorted system has a loss of about $1$\,dB compared to the ideal PA, at least for the SNR values considered in the figure. On the other hand, with the multi-carrier waveform a minimum $\rm BER \approx 10^{-4}$ can be achieved with an $\rm{SNR} \approx 23.5$\,dB. For $\rm{SNR} > 23.5$\,dB, the polynomial predistorter is unable to compensate for the \gls{pa} distortions, and the BER increases. 

Similar trends can be observed in Fig.~\ref{fig:ber_vs_SNR_N=8}, which illustrates the performance of the polynomial system with $N_{A}$ = $N_{\theta}= 8$. It is seen that, increasing the polynomial order improves performance significantly, achieving a near-ideal linearization for both single-carrier and multi-carrier waveforms, in a wide range of SNR values.

%\begin{figure}[]
%	\centering
%	\subfloat[]{\includegraphics[width=0.5\columnwidth]{Figures/Simulation_Results/ber_N=4.eps}\label{fig:ber_vs_SNR_N=4}}
	%\hfil
%	\subfloat[]{\includegraphics[width=0.5\columnwidth]{Figures/Simulation_Results/ber_N=8_.eps}\label{fig:ber_vs_SNR_N=8}}
%	\caption{BER vs SNR}\label{fig:BERvsSNR}
%\end{figure}

\begin{figure}[h]
	\centering
	\subfloat[]{\includegraphics[width=0.5\columnwidth, height=3.2cm]{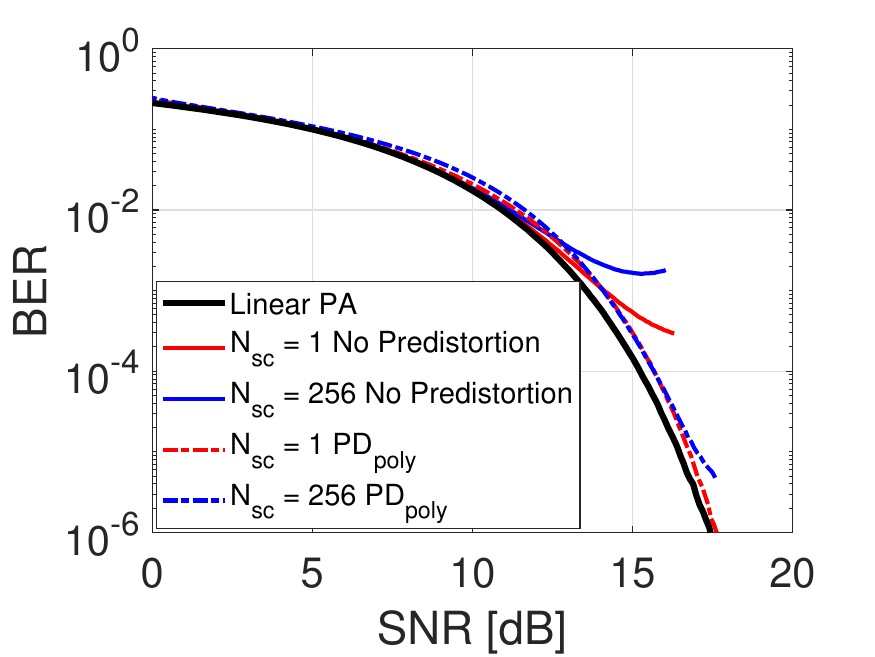}\label{fig:ber_vs_SNR_N=4_16QAM}}
	%\hfil
	\subfloat[]{\includegraphics[width=0.5\columnwidth, height=3.2cm]{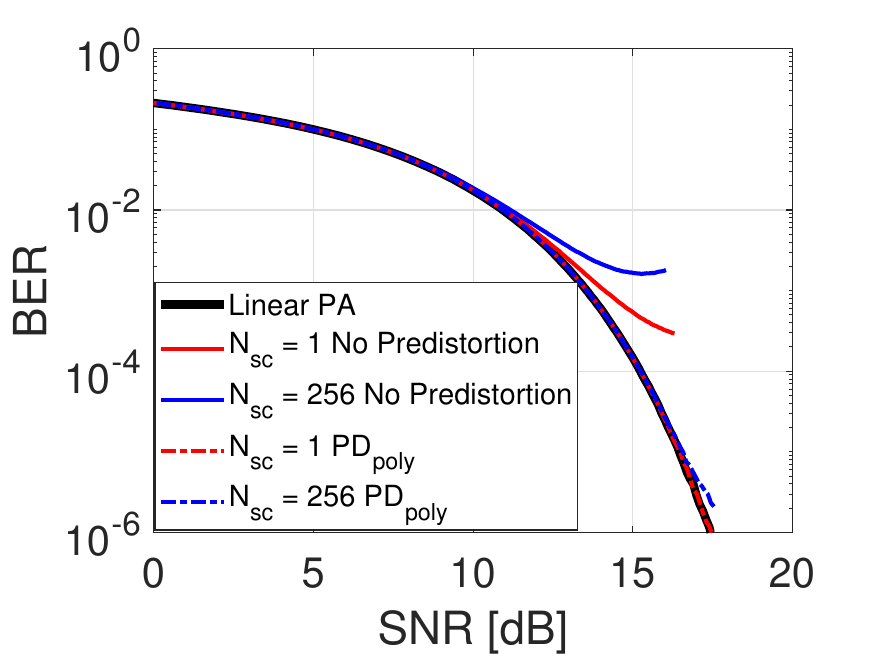}\label{fig:ber_vs_SNR_N=8_16QAM}}
	\caption{BER vs SNR performance, 16 QAM. (a) $N_A=4,N_\theta=4$, (b) $N_A=8,N_\theta=8$}\label{fig:BERvsSNR_16QAM}
\end{figure}

The same conclusions can be drawn from the results in Fig.~\ref{fig:BERvsSNR_16QAM} which have been obtained with a $16$-QAM constellation. In this case, the distance between trasnmitter and receiver is $d = 70$ m. As can be seen, the predistortion algorithm achieves good performance  with both $N_A=N_\theta=4$ and $N_A=N_\theta=8$, and provides substantial improvements compared to the case of no compensation of the PA non-linearities. Also, marginal differences can be observed between the bit error rates in the single-carrier case and in the multi-carrier case. 

Simulation results -- that have not been reported to avoid overloading the figures -- indicate that reducing the number of subcarriers (in particular, passing from  $N_{sc} = 256$ to $N_{sc} = 16$) has a negligible effect on BER performance, as expected from the EVM behavior illustrated in Fig.~\ref{fig:evm_vs_nsc_ibo0} and  Fig.~\ref{fig:evm_vs_nsc_ibo10}.

\section{Discussion and Concluding Remarks}
In this paper, the characterization of the behavior of a \gls{pa} in the 300\,GHz band was carried out, where the dependence of the AM-AM and AM-PM behaviors on the operating frequency was captured. Various models have been derived and verified using results from a validation set-up, where a relative accuracy was noted in modeling the errors resulting from distortions in the compression region. The parameters of different AM-AM and AM-PM distortion behavioral models are provided for different operating frequencies, namely Saleh, Ghorbani, polynomial, and the modified Rapp models. Based on the derived models, pre-distortion was applied on single-carrier and multi-carrier (\gls{ofdm}) waveforms, where performance analysis revealed the impact of several pre-distortion design parameters, such as the order of the polynomial approximation of the inverse AM-AM and AM-PM distortion functions and IBO. The analysis revealed a significant dependence of the error performance when moving form a single-carrier to a multi-carrier set-up, while the error performance did not show any relevant dependency on the number of subcarriers, regardless of the application of pre-distortion. Moreover, the analysis showed that pre-distortion significantly improves the error performance, thus providing the option of reliably adopting multi-carrier waveforms at 300\,GHz.

%\section*{Acknowledgment}
%This work has been performed in part in the framework of the HORIZON-JU-SNS-2022 project TIMES, co-funded by the European Union. Views and opinions expressed are however those of the author(s) only and do not necessarily reflect those of the European Union. 

\bibliographystyle{ieeetr}
\bibliography{main.bib}

\end{document}